\documentclass[useAMS,usenatbib]{mn2e}
\usepackage{mathrsfs} 
\usepackage[dvipdfmx]{graphicx}
\usepackage{mediabb}
\usepackage{color}  
 
\def\kms{km s$^{-1}$}
\def\vlsr{V_{\rm lsr}}
\def\vrot{V_{\rm rot}}
 
\def\vexpa{V_{\rm expa}}
\def\vrot{V_{\rm rot}}
\def\voval{V_{\rm oval}}
\def\Msun{M_\odot}
\def\~{$\sim$} 
\def\deg{^\circ} 
\def\Tb{T_{\rm B}} 
\def\acmz{a_{\rm CMZ}} \def\aemr{a_{\rm EMR}}

\def\Hcc{{\rm H\ cm^{-3}}}

\title{The 200-pc Molecular Cylinder in the Galactic Centre} 
\author[Yoshiaki SOFUE]{Yoshiaki \textsc{Sofue}$^{1}$\thanks{E-mail: sofue@ioa.s.u-tokyo.ac.jp}\\
$^1$Insitute of Astronomy, The University of Tokyo, Mitaka, Tokyo 181-0015, Japan }   

\begin{document} 
\date{ } 

\maketitle  

\begin{abstract} 
{
Analyzing the 3D structure of the molecular gas distribution in the central 200 pc region of the Galaxy, we show that the expanding molecular ring (EMR, also known as the parallelogram) and the central molecular zone (CMZ) exhibit quite different distributions and kinematics. The EMR composes a bipolar vertical cylinder with the total length as long as $\sim 170$ pc and shows large non-circular velocities. On the contrary, the CMZ is distributed in a nearly rigid-body rotating ring and arms tightly concentrated near the galactic plane with full thickness less than $\sim 30$ pc. Furthermore, the mass and density ratios of the EMR to CMZ are as small as $\sim 0.13$ and 0.04, respectively. We discuss the origins of the EMR and CMZ based on the bar and explosion models. We suggest that the EMR's large vertical extent can be explained by a cylindrical shock wave model driven by an explosive activity in the Galactic Centre. 
}
\end{abstract}

\section{Introduction}
 
The main structure of the molecular gas in the Galactic Center (GC), the central molecular zone (CMZ), is distributed near the galactic plane at $|l| \le \sim 1-2\deg$ (150 pc) and $|b|\le 0\deg.1$ (15 pc). Kinematics shown by longitude-velocity (LV) diagrams indicates nearly rigid-body rotation (Heiligman 1987; Sofue 1995a; Henshaw et al. 2016). Henshaw et al. (2016) analyzed the detailed kinematics and spatial distribution of the high-density gas, and showed that the CMZ is composed by several arms with projected lengths $\sim 100-250$ pc. The star forming regions, Sgr B and C, are associated with molecular complexes in the CMZ at the leading ends of the two rotating arms named Arm I and II (Sofue 1995a; Sawada et al. 2004).

Besides the CMZ, a more extended molecular gas structure is known as the 200-pc expanding molecular ring (EMR), while its brightness is an order of magnitude fainter than the CMZ. The EMR was recognized as a tilted ellipse on the LV diagram with large non-circular velocities (Kaifu et al. 1972, 1974; Scoville 1972; Oort 1977). By a wide-area CO-line survey, the EMR was shown to be vertically extended for more than $b\sim \pm 0\deg.4$ (Bally et al. 1987). It was shown that the EMR is even more extended, reaching to $b \sim \pm 0\deg.6$ ($\pm 84$ pc), by a larger-scale, higher-resolution CO-line survey using the Nobeyama 45-m telescope (Oka et al. 1998). 

Although the CMZ and EMR appear to be located close to each other in the GC, they exhibit quite different kinematics and geometry. In this paper, we investigate their 3D structures and mutual relations. We revisit the vertically extended cylindrical structure of the EMR following the idea presented by Sofue (1995b), and call the structure the EMC (expanding molecular cylinder). Throughout the paper, we assume that the distance to the GC is $R_0\simeq 8$ kpc (Honma et al. 2015). 

\section{3D Structure of the EMR}
 
\subsection{Data}
We use the high-resolution and wide-area $(l,b,v)$ data cube of the $^{12}$CO ($J=1-0$) $\lambda$ 2.63-mm line emission observed with the Nobeyama 45-m Telescope by Oka et al. (1998). The telescope had a beam width of FWHM of $15''$, and the observational grid interval was $30''$, yielding an effective spatial resolution of $34''$ (13 pc). The velocity interval of the data cube was 5 \kms. The data cover a wide area around the GC, and we analyze the EMR/CMZ region at $-1\deg.5 \le l \le 1\deg.8$ ($-200$ to 250 pc) and $-0\deg.6 \le b +0\deg.6$ (-84 to $+84$ pc). 

\subsection{Vertical extension}

\begin{figure} 
\begin{center} 
\includegraphics[width=0.7\linewidth]{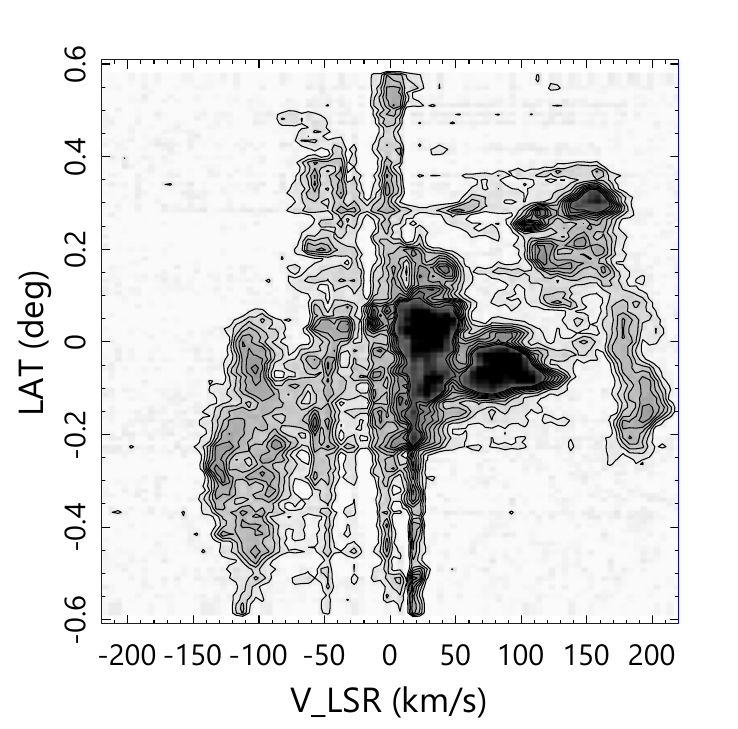} \\ 
\includegraphics[width=0.7\linewidth]{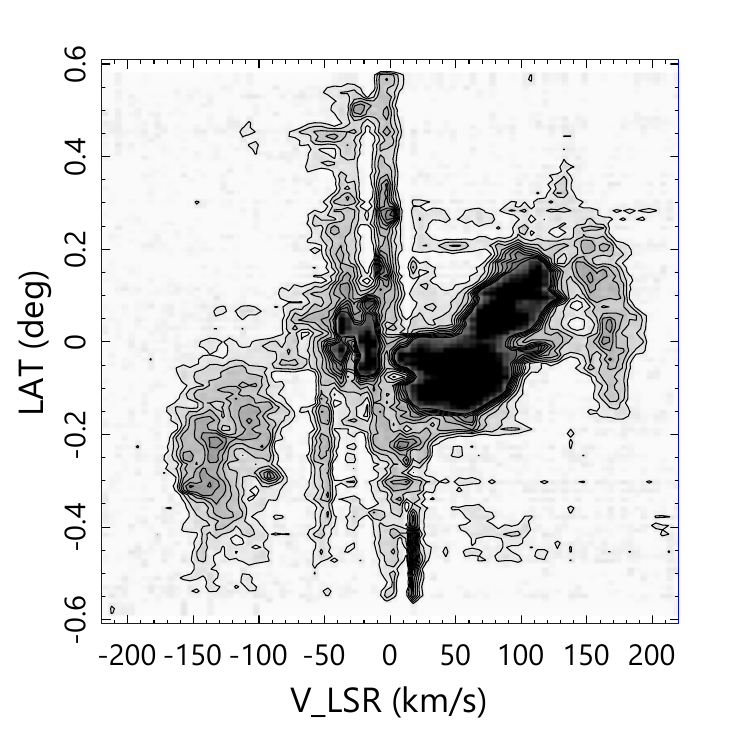}  \\
\includegraphics[width=0.7\linewidth]{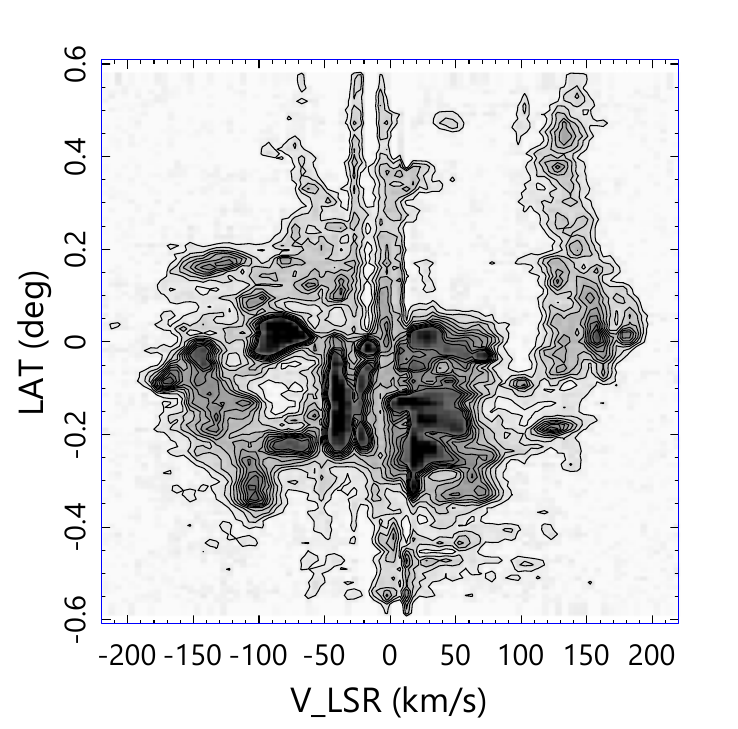}  \\ 
\vskip -171mm
{\bf ~~~~~~~~~~~ Local ~~~~~~~~~~~~~~~~~~~~}\\
{\bf ~~~~~~~~~~~~~~~~~~~~~~~~~~~~~~ EMR$+$}\\
\vskip 16mm
{\bf ~~~~~~~~~~~~~~~~~~~~~~~~\textcolor{white}{CMZ} ~~~~~~~~}\\
\vskip 12mm
{\bf EMR$-$ ~~~~~~~~~~~~~~~~~~~~~~~~~~~ }\\
\vskip 130mm
\end{center}
\caption{BV (latitude-velocity) diagram of the CO line intensity in 5 \kms velocity bins at 
 $l=0\deg.4$ (top), $0\deg$ (middle) and  
$-0\deg.4$ (bottom). Contour interval is 2 K \kms starting at 2K \kms (or 0.4 K in brightness temperature). Note the extended ridges at $\vlsr \sim \pm 100-150$ \kms, showing the vertical extension of the EMR. Long stripes at $\vlsr\sim -50$ to $\sim 20$ \kms are foreground disk emission and absorption.}
 \label{BVD}  
\end{figure}  
 
In order to investigate the vertical structure of the EMR, we show latitude-velocity (BV) diagrams of the CO line emission in the Galactic Centre sliced at $l=-0\deg.4,\ 0\deg,$ and $+0\deg.4$ in figure \ref{BVD}. 
Ignoring the vertical stripes caused by the foreground disk emission and absorption, the diagrams exhibit two major components. The most massive component is the dense and concentrated features at $|\vlsr|\le \sim 50$ \kms and $|b|\le \sim 0\deg.1-0\deg.2$, which represent the CMZ. Another component is the vertically extended ridge-like features at $|\vlsr|>\sim 100$ \kms, representing the EMR. The ridges in the BV plane extend from $b\sim -0\deg.6$ to $\sim +0\deg.2$ at negative velocities, and from $\sim -0\deg.2$ to $\sim +0\deg.6$ at positive velocities. 

The top and bottom ends of the vertical ridges reach the observed latitudinal edges at $0\deg.6$ (84 pc), and seem to be extending further to higher latitudes. This fact indicates that the EMR is a structure vertically extended for more than $\sim 84$ pc from the galactic plane, and hence the total length in the $Z$ direction is greater than $\sim 170$ pc, almost comparable to its radius.
 
The EMR is known to be inclined from the galactic plane (Burton and Liszt 1978; Liszt and Burton 1978, 1980). In the present data, EMR appears asymmetric in the BV plane in the sense that the positive-velocity side is displaced to positive latitudes, and vice versa in the negative velocities. If the EMR is an expanding ring, or alternatively an ellipse along an oval orbit, the ring or the orbit is significantly inclined from the line of sight. 

\begin{figure} 
\begin{center}  
(a)\includegraphics[width=0.9\linewidth]{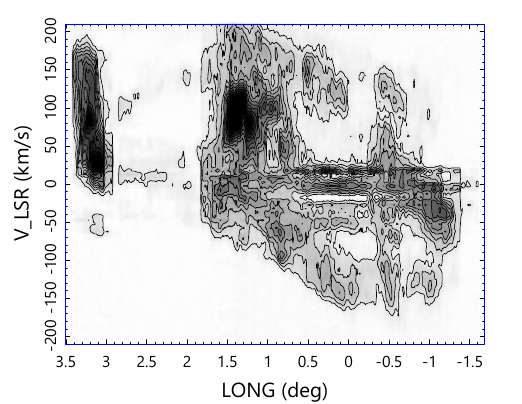} \\ 
(b) \includegraphics[height=0.34\linewidth]{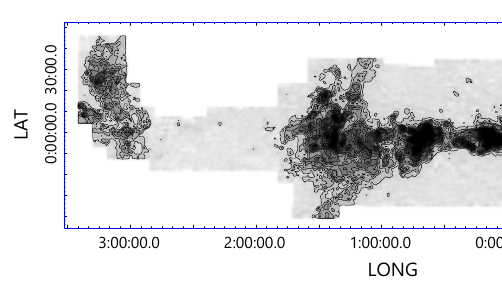} 
\hskip -1.3mm
\includegraphics[height=0.34\linewidth]{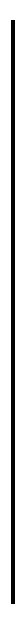}   
\hskip -1.3mm
\includegraphics[height=0.34\linewidth]{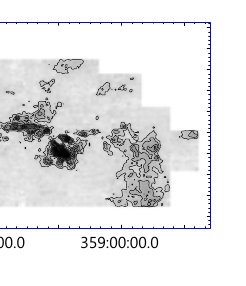}  \\
 \hskip 28mm $<$------CMZ------$>$\\
 \hskip 28mm $<$---------EMR------$>$\\ 
\end{center}
\caption{
(a) LV diagram averaged for out-of-plane EMR/EMC at $|b|\ge 0\deg.2$. Contour interval is 0.5 K \kms.  The CMZ near the galactic plane, mostly at $|b|<0\deg.2$, does not show up here. The horizontal stripes are due to foreground spiral arms.
(b) Mosaic intensity maps at $\vlsr=+67.5$ \kms for $l>0\deg$, and at -67.5 \kms for $l<0\deg$. Note the large vertical extension of the EMR/EMC at $l\sim \pm 11\deg.2$, suggesting further extension beyond $\pm 0\deg.6$ ($\pm 84$ pc).  
}
 \label{LVandLB}  
\end{figure} 

\begin{figure} 
\begin{center}   
(a)\includegraphics[width=0.8\linewidth]{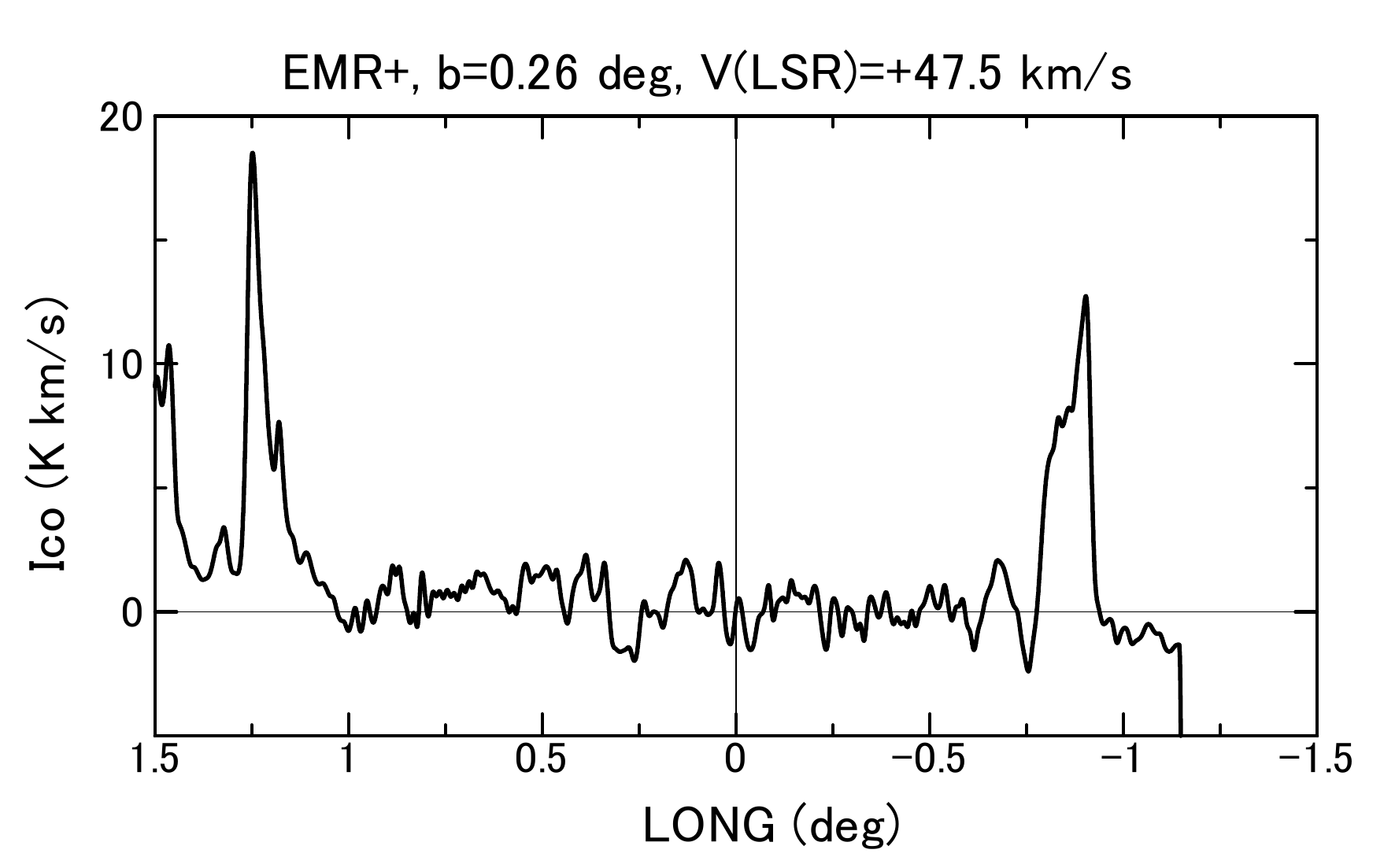} \\
(b)\includegraphics[width=0.8\linewidth]{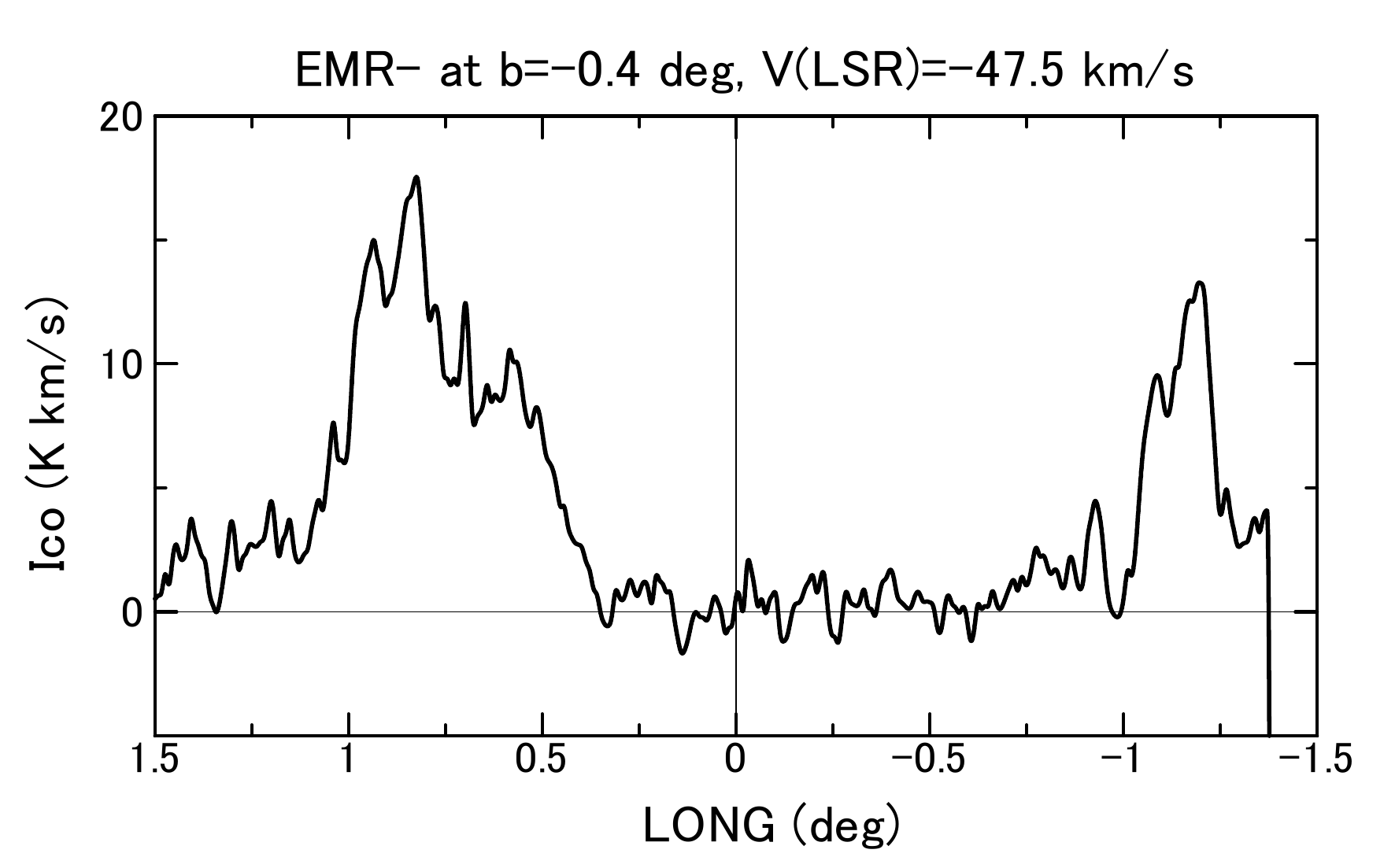}  
\end{center}
\caption{ 
(a) Horizontal cross section of the integrated intensity of the EMR+ at $b=+0\deg.26$, $\vlsr=+47.5$ \kms. 
(b) Same of the EMR- at $b=-0\deg.4$, $\vlsr=-47.5$ \kms. 
  }
 \label{Horizoncut}  
\end{figure} 

Figure \ref{LVandLB}(a) shows an LV diagram averaged at $|b|\ge 0\deg.2$ in order to show the elliptical shape of the LV ridge for the high-latitude gases in the EMR. Panel (b) shows mosaic intensity channel maps at $\vlsr=+67.5$ and $-67.5$ \kms corresponding to the positive- and negative-tangential velocities of the EMR ellipse in panel (a), showing the projected distributions of the EMR gas on the sky sliced at the velocities. The gases are largely extended in the vertical directions at $l=1\deg.5$ and $-1\deg.2$, indicating that the EMR is an elongated structure in the vertical direction over $0\deg.6$, even beyond the observed $b$ edges. This means that the EMR is extending for more than $\sim \pm 84$ pc at $l\sim 1\deg.5$ and toward negative latitude for more than $\sim -84$ pc. So, the total vertical extent of the EMR is larger than $\sim 170$ pc. These extents are consistent with those estimated from the BV diagrams in figure \ref{BVD}.

Figure \ref{Horizoncut} shows horizontal cross sections of the EMR at a constant latitude of $b=+0\deg.26$ in the channel map at $\vlsr=+47.5$ \kms\ (EMR+), and at $b=-0\deg.4$, $\vlsr=-47.5$ \kms\ (EMR-). The EMR+ appears as the two-horn like intensity peaks at $l\sim 1\deg.2$ and $\sim -0\deg.9$, and the EMR- at $l\sim 0\deg.8$ and $\sim -1\deg.2$. It is stressed that the CMZ does not show up in the figure at all, indicating that it is a tightly concentrated, apparently thin structure near the galactic plane in contrast to the vertically extended EMR ridges.

\subsection{Measurement of the vertical extents}

\begin{figure} 
\begin{center}  
(a) \includegraphics[width=0.8\linewidth]{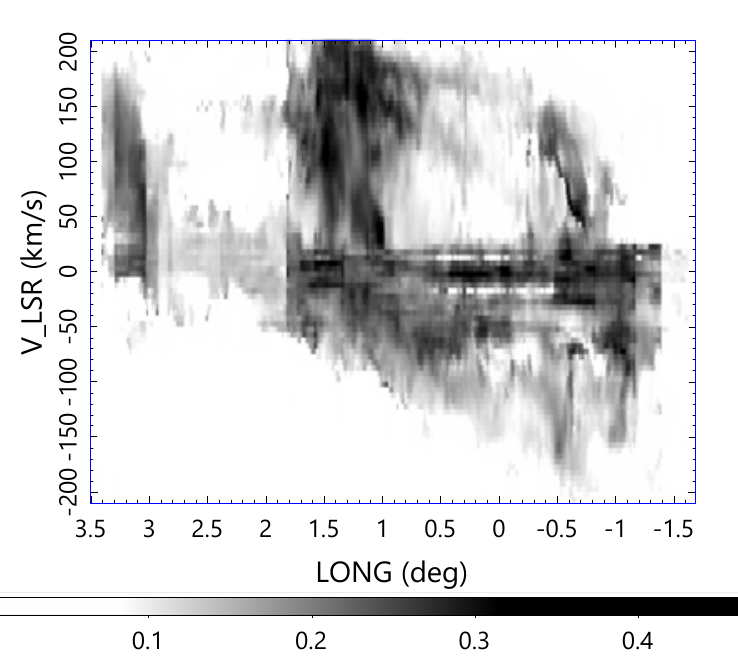} \\
(b) \includegraphics[width=0.8\linewidth]{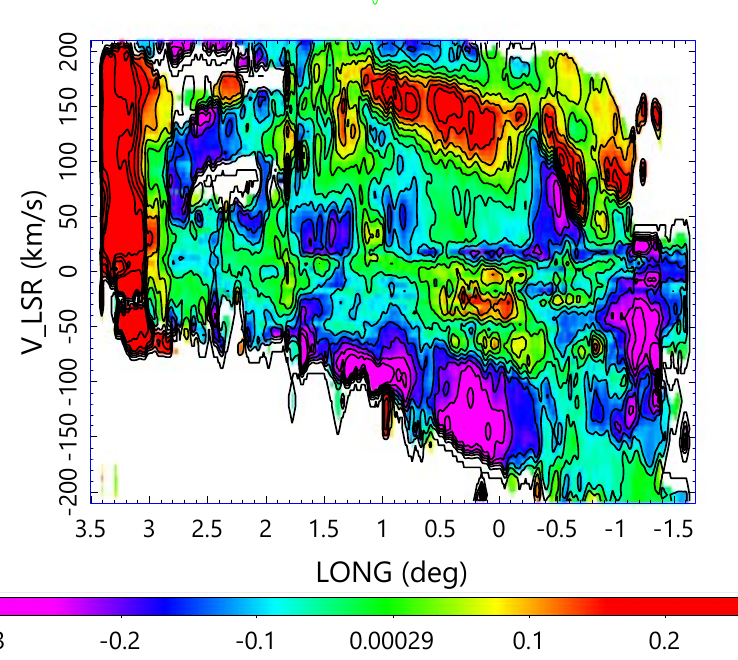}  \\ 
\end{center}
\caption{ 
(a) Latitudinal thickness $\Delta b$ in the LV plane, showing larger thickness of the EMR than the other regions including CMZ. 
(b) Mean latitudinal displacement, showing the inclined EMR. }
 \label{M12}  
\end{figure} 

We then measure the vertical length and the displacement of the EMR more quantitatively using the CO-line data cube.
Figure \ref{M12} shows moment maps of the surveyed area in the CO line by Oka et al. (1998). Diagrams (a) and (b) show moment 2 and moment 1 maps of the $(b,l,v)$ cube with respect to latitude $b$, respectively. The moment 2 map shows the distribution of dispersion in the galactic latitude $\Delta b$, representing the vertical extent distribution, and moment 1 map shows the mean height ${\bar b}$.

\begin{figure} 
\begin{center} 
\includegraphics[width=0.8\linewidth]{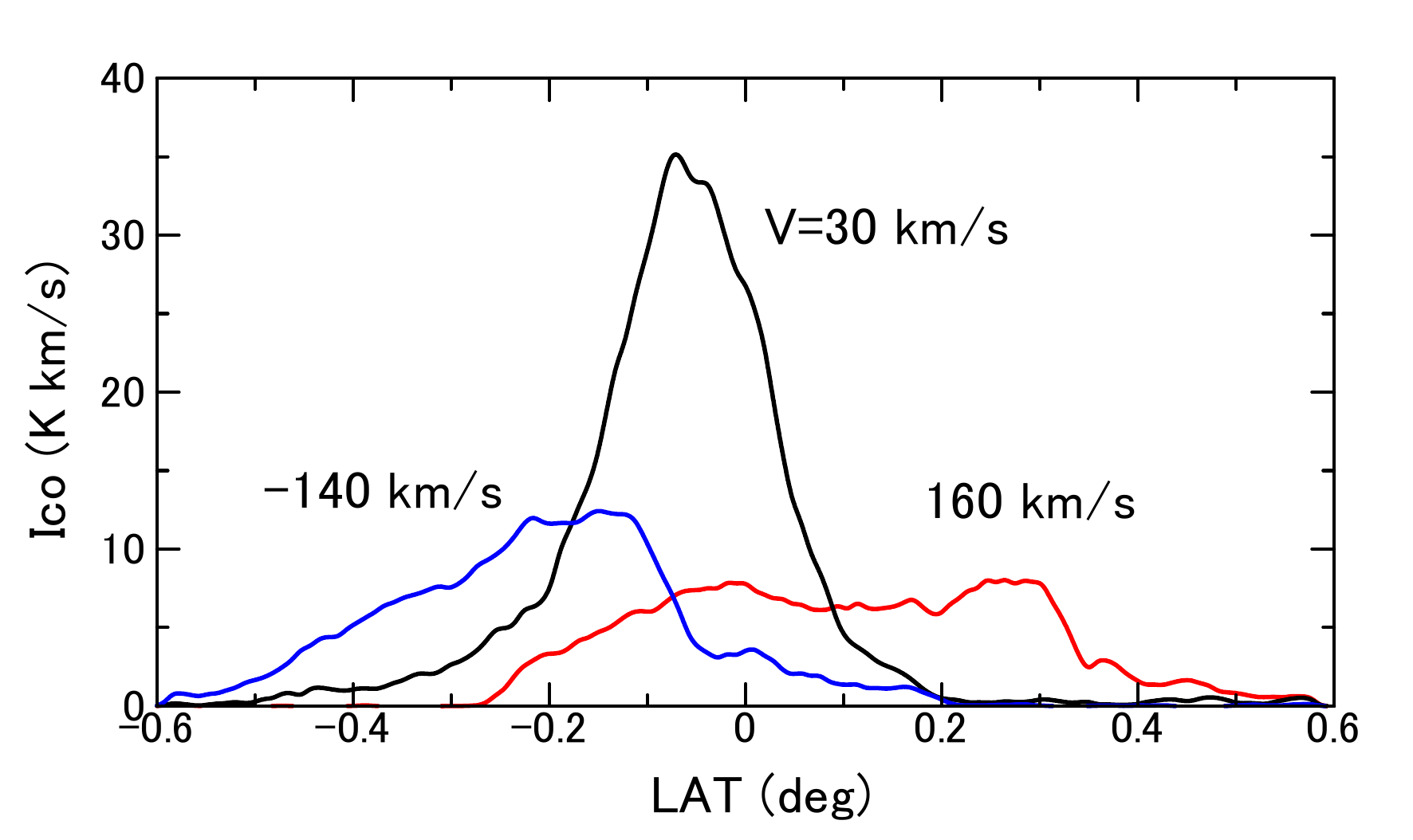} 
\end{center}
\caption{Latitudinal intensity variation in the BV diagram at $\vlsr \sim -140,\ 30$, and 160 \kms corresponding EMR$-$, CMZ, and EMR+, respectively. The vertical scale is the intensity integrated in 5 \kms velocity bins. Note the wider spread and larger displacement from the galactic plane of EMR than CMZ.}
 \label{BVcut}  
\end{figure}

The EMR is traced by the grey elliptical region on the moment 2 map, showing a vertical extent $\Delta b$ as large as $\sim 0\deg.2$ ($\sim 30$ pc) on average on the ellipse. On the contrary, the CMZ region exhibits much smaller extent of $\Delta b \sim 0\deg.1$ (15 pc). Figure \ref{BVcut} shows latitudinal intensity distributions at representative $\vlsr$ at $l=0\deg$, showing CMZ $(\vlsr=30$ \kms) and positive- and negative-velocity EMR (160 and $-140$ \kms). The CMZ is shown to have approximately a Gaussian $b$ profile with the vertical extent of $\Delta b\sim 0\deg.1$. On the other hand, the EMR shows a broad and rather plateau-like profile with much large vertical extent of $\Delta b\sim 0\deg.3-0\deg.4\ (40-55)$ pc. In fact, the maximum extent of the $b$ profile at $\vlsr=160$ \kms amounts to $\sim 0\deg.6$ or $\sim 84$ pc.

\begin{figure} 
\begin{center} 
\hskip -2mm(a)\includegraphics[width=0.82\linewidth,height=0.44\linewidth]{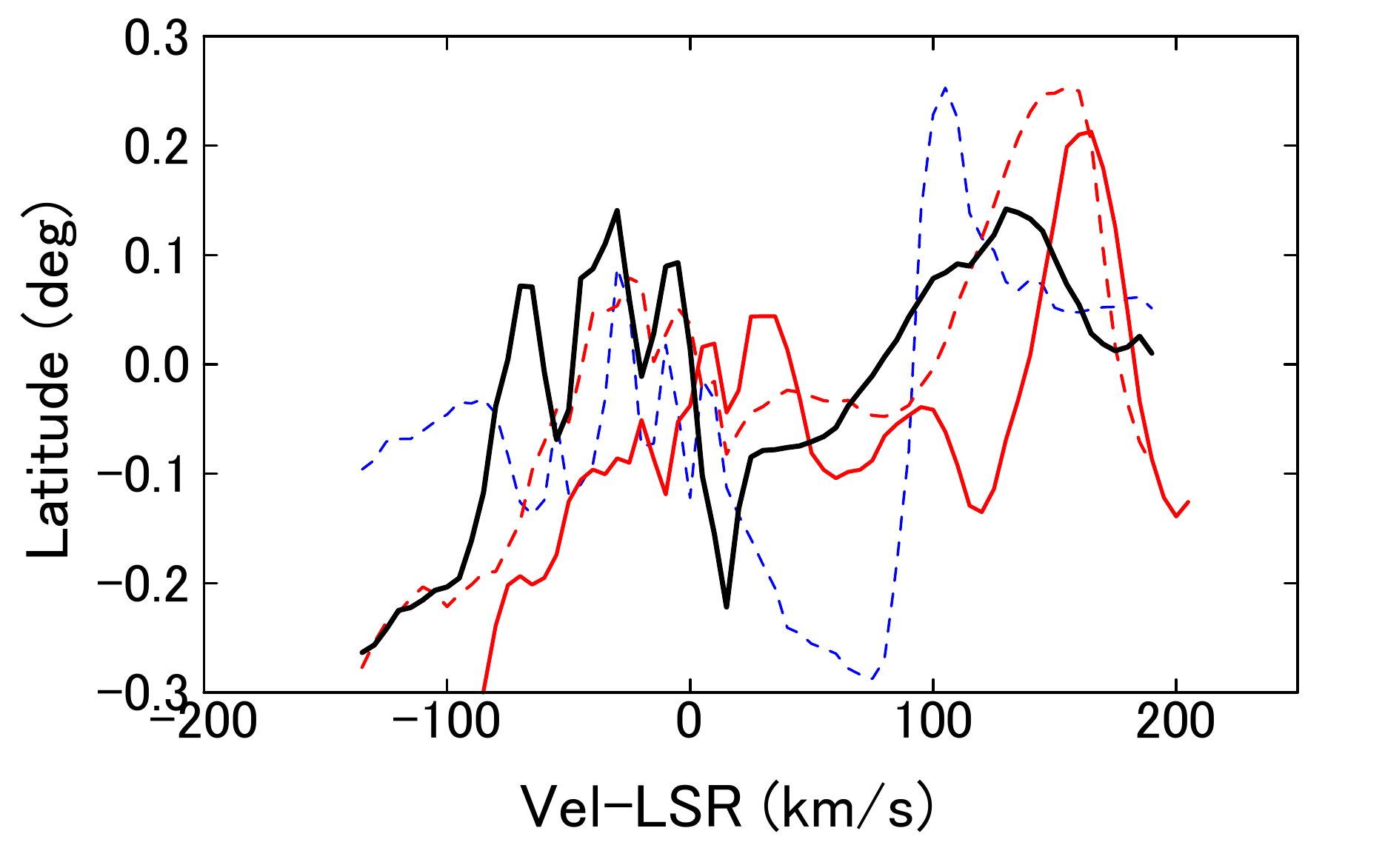}  \\
(b)\includegraphics[width=0.8\linewidth]{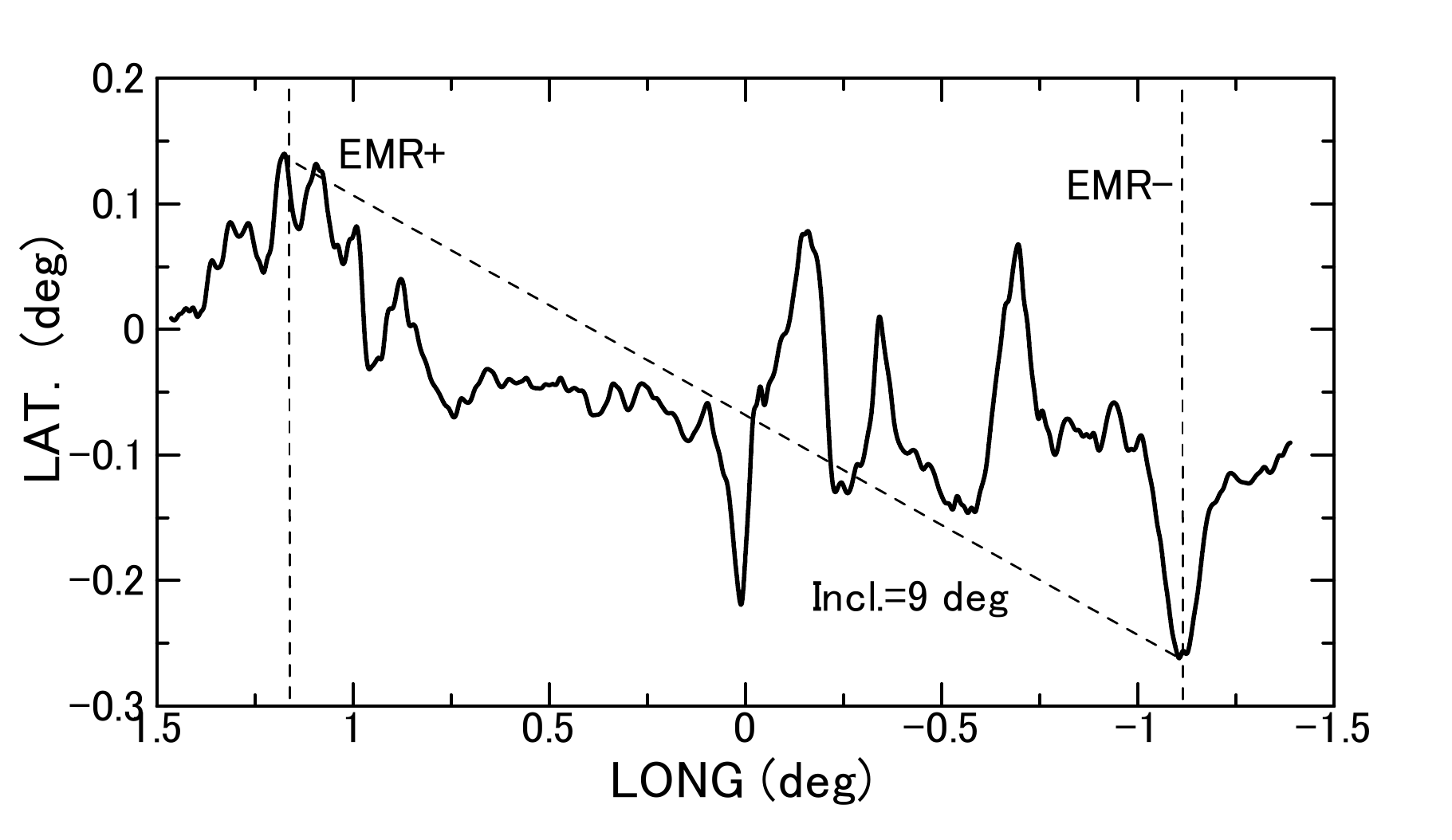}  
\end{center}
\caption{(a) Variation of the mean $b$ displacement against $\vlsr$ at $l=-0\deg.5$ (blue dash), $0\deg$ (black solid), $+0\deg.5$ (red dash) and 1 degrees (red), showing the tilt of EMR on the line of sight.
(b) Mean $b$ displacement against $l$ along the major axis of the LV ellipse of EMR. The EMR+ and EMR- are inclined by $9\deg$ from the galactic plane.}
 \label{M1cut}  
\end{figure} 

 The tilt of the mean latitudinal heights ${\bar b}$ in the moment 1 map in figure \ref{M12}(b) is also clearly recognized by cross sections at fixed longitudes shown in figure \ref{M1cut}(a), where the mean $b$ displacement is plotted against $\vlsr$ at $l=-0\deg.5,\ 0,\ +\deg.5$ and $+1\deg$. An averaged tilt of the plots is approximately measured to be $d{\bar b} /d\vlsr \simeq 0\deg.2/200$ \kms, or $\sim 28$ pc/200 \kms. If we assume that the line-of-sight distance of the two peaks is $\sim \pm 200$ pc, the tilt angle against the line of sight is $\sim 8\deg$.
 
 Figure \ref{M1cut}(b) shows the mean $b$ displacement along the major axis of the LV ellipse, showing that the mean heights of the eastern and western edges are systematically displaced upward by ${\bar b} \sim 0\deg.15$, and downward by $\sim -0\deg25$ from the galactic plane. This indicates a tilt of the EMC's equator about $9\deg$ from the galactic plane on the sky. Combining this with the tilt from the line of sight, the absolute tilt angle of the EMC's equator is estimated to be $\sim 12\deg$.

\subsection{3D Cylindrical structure}  

\begin{figure} 
\begin{center}  
\includegraphics[width=0.8\linewidth]{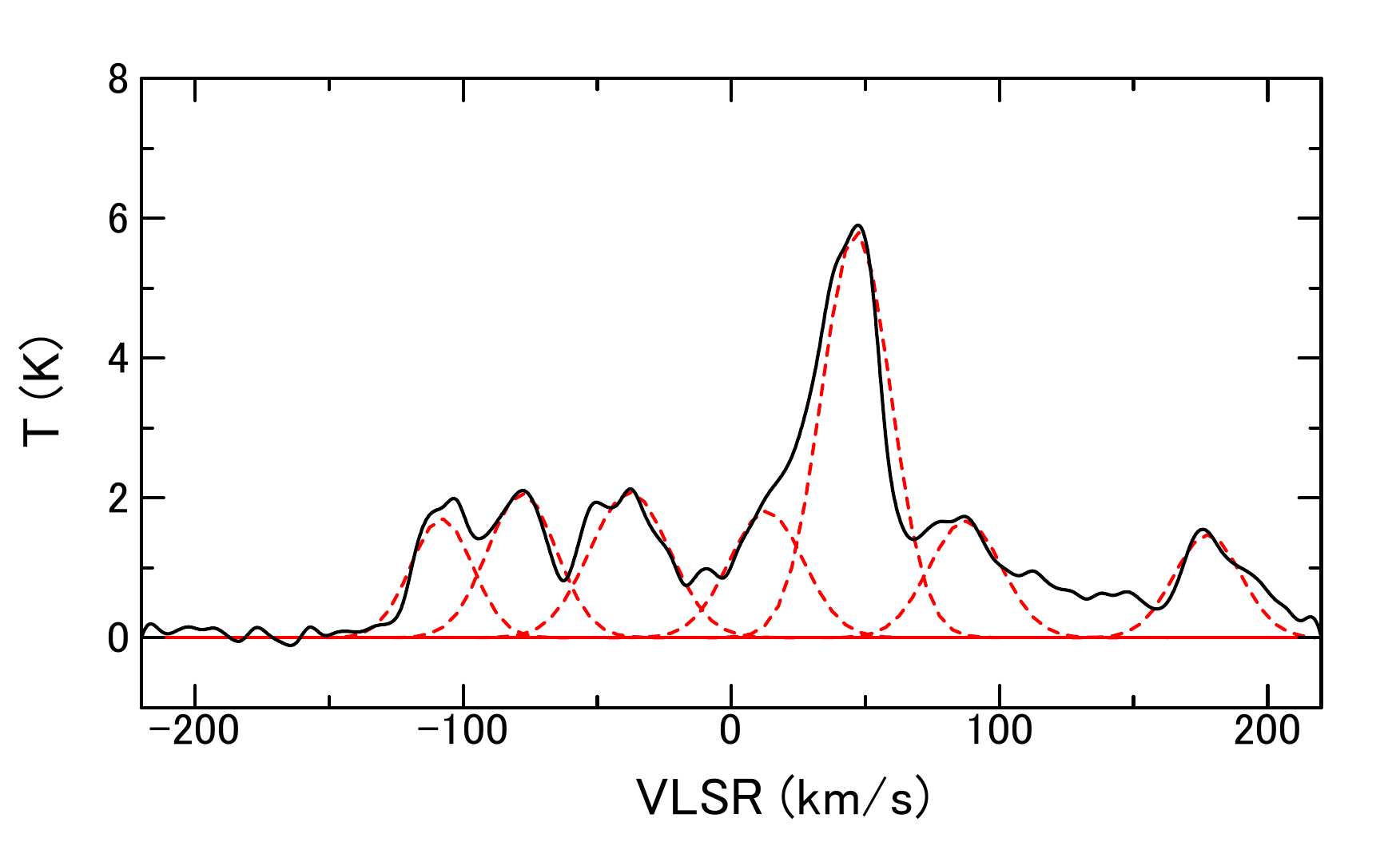} 
\end{center}
\caption{Line spectrum at $(l,b)=0\deg.7, -0\deg.3$ and Gaussian fitting result for velocity component deconvolution. }
 \label{spec}  
\end{figure}  

In order to understand the 3D relationship between the EMR and CMZ, we construct 3D distribution plots of peak intensity positions in the $(l,b,v)$ place of the CO line emission, following the method employed by Henshaw et al. (2016).
At each $(l,b)$ data point in the CO data cube by Oka et al. (1988), we decomposed the line spectrum into Gaussian components. 
Figure \ref{spec} shows an example of the decomposition of the spectrum at $(l,b)=(0\deg.7, -0\deg.3)$. Thereby, only line components with peak intensities greater than 1 K have been adopted, but components with lower intensities are not picked up, since we aim at obtaining 3D distribution of the velocity components, but not the intensity distribution. 

\begin{figure*} 
\begin{center} 
\includegraphics[width=0.55\linewidth]{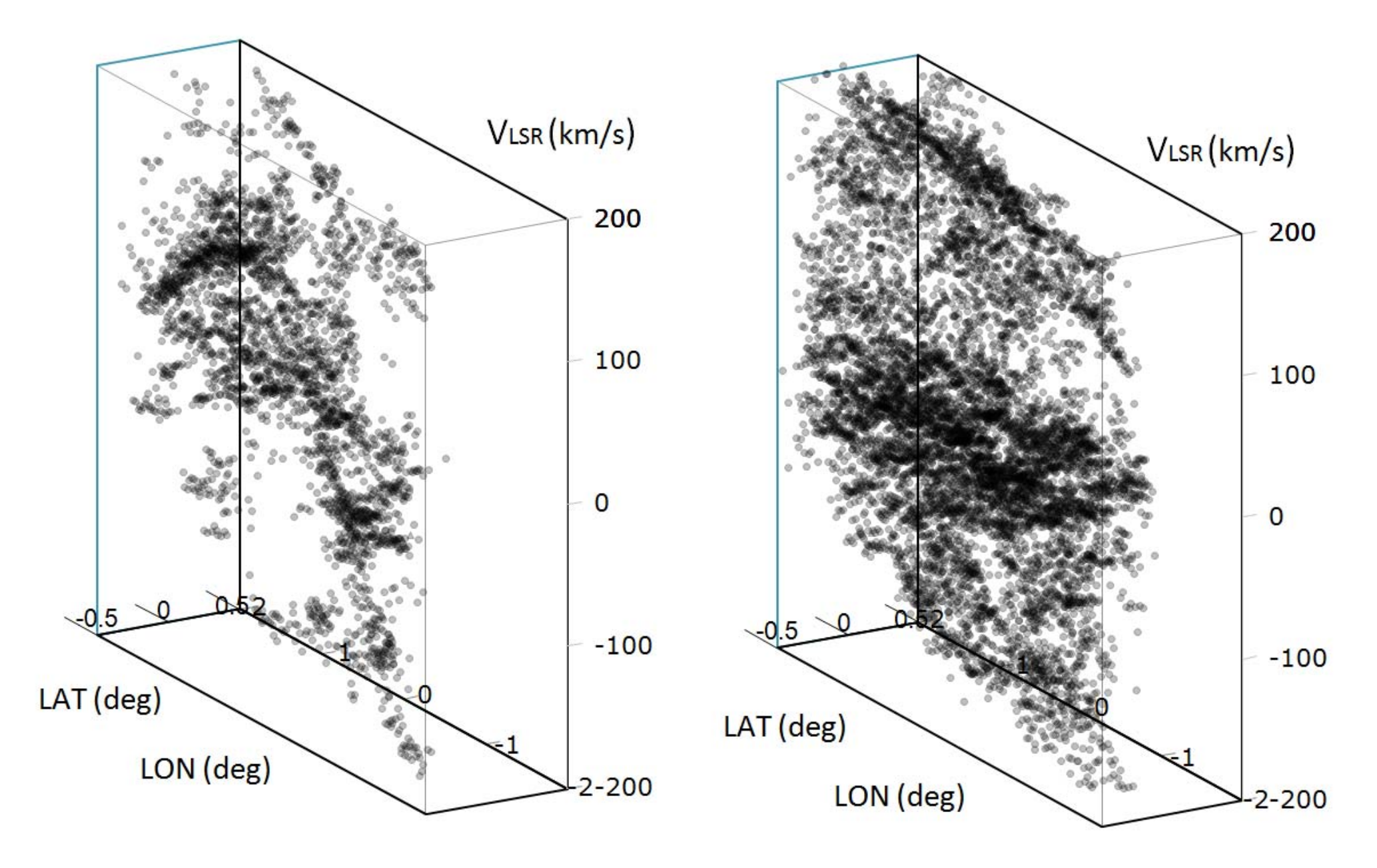}
\includegraphics[width=0.27\linewidth]{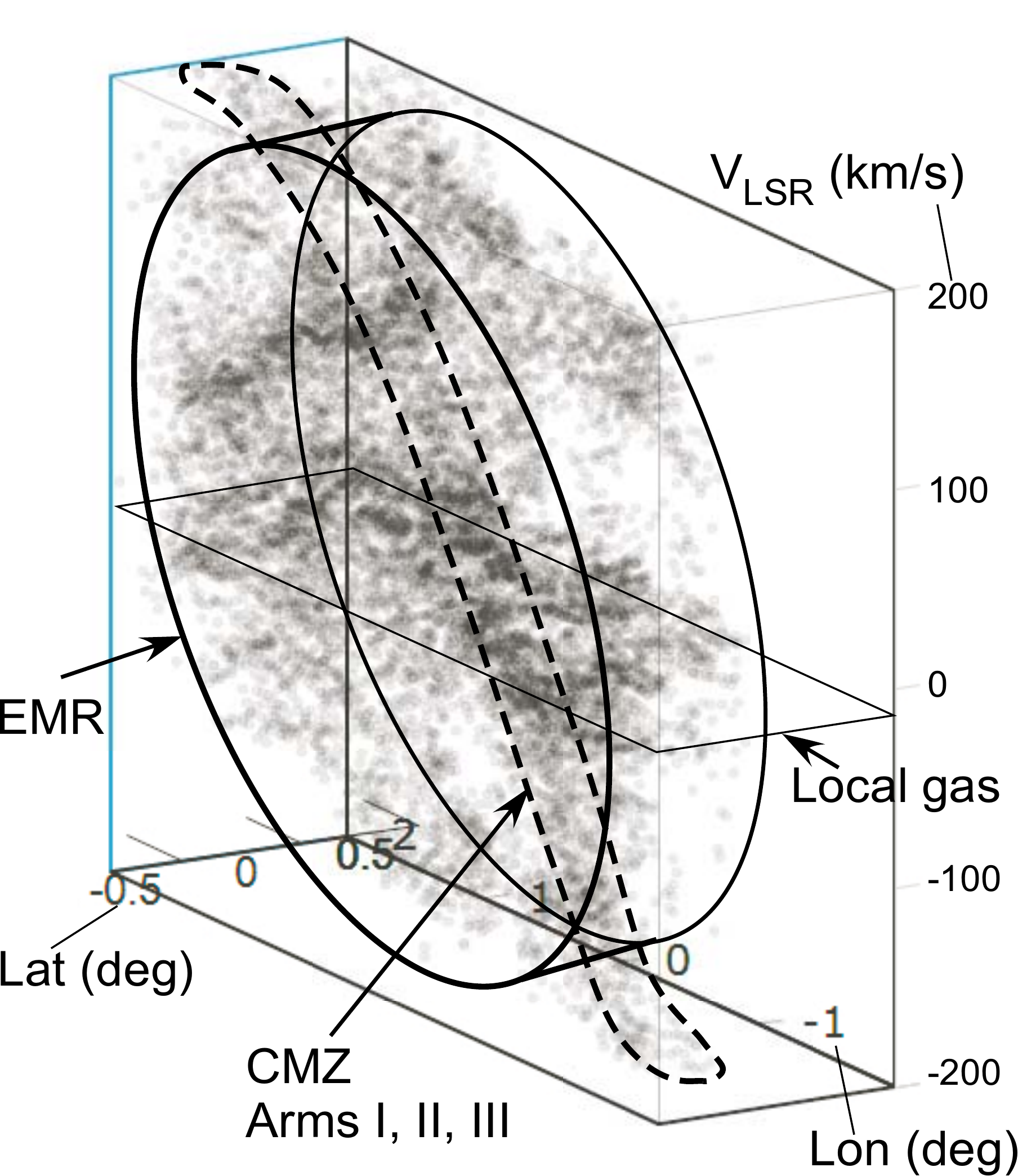} 
\\
(a) ~~~~~~~~~~~~~~~~~~~~~~~~~~~~~~~~~~~~~~~~~~~~~~~~~~~~~~(b)~~~~~~~~~~~~~~~~~~~~~~~~~~~~~~~~~~~~~~~~~~~~~~~~~(c)\\
\end{center}
\caption{(a) 3D plot of velocity components with peak $\Tb >3$ K, showing the distribution of high-density gas in the CMZ. The rigid-body like LV ridge shows up clearly, representing rotating ring or arms. (b) Same, but lower temperature components with $\Tb \le 3$ K, showing extended components and EMR. (c) Schematic illustration of the EMR and CMZ. }
 \label{3Db}  
\end{figure*}   

Using the decomposed result, we plot the peak positions in the ($l,\ b,\ \vlsr$) space as a 3D spatial representation. Figure \ref{3Db}(a) shows the thus obtained 3D plots of the line peak positions having high brightness of $\Tb > 3$ K at viewing angle of $\phi=60\deg$ and rotation angle of $\psi=20\deg$ around the longitude axis. The figure shows 3D LV behavior of higher-density molecular clouds, which compose the CMZ. The figure may resemble the 3D plots obtained by Henshaw et al. (2016) for the molecular lines representing high-density gases.
The figure exhibits the general LV property of the CMZ showing a rigid-body like tilted LV ridge running across $(l,\vlsr)\sim(0,0)$, indicating a rotating ring or arm structure. 

Figure \ref{3Db}(b) shows the same, but represents more extended, less bright components with $\Tb \le 3$ K. The plots are spread both in the $b$ and $\vlsr$ directions. These low temperature components represent the distribution of lower-density gas, mostly composing the EMR. Note that fore-/and background galactic features appear as the plane-like distributions at low velocities. 
It should be mentioned that the figures show the number distributions of velocity components, but do not show the gas density. This is the reason why a larger number of components was found in the EMR than in the tightly concentrated CMZ near the Galactic plane. 

Comparison of panels (a) and (b) in figure \ref{3Db} indicates that the EMR is a largely extended structure in the $b$ direction, composing a cylindrical structure, much thicker than the CMZ. Hereafter, we call the structure the expanding molecular cylinder (EMC). We illustrate the relationship of the CMZ and EMR in the 3D plot in panel (c) of figure \ref{3Db}.

\subsection{Relative Location of EMR to Arms I and II of CMZ}

Sawada et al. (2004) derived the face-on geometry of Arms I, II and EMC by analyzing the 18-cm OH line absorption against the circum-nuclear continuum emission. From the optical depth map, they showed that Sgr B and C molecular complexes are located at $y\sim -30$ and $-90$ pc, respectively, both inside the EMC. We apply this method to the present data in order to locate the EMC relative to the CMZ with Arms I and II.

\begin{figure} 
\begin{center} 
(a)\includegraphics[width=0.7\linewidth]{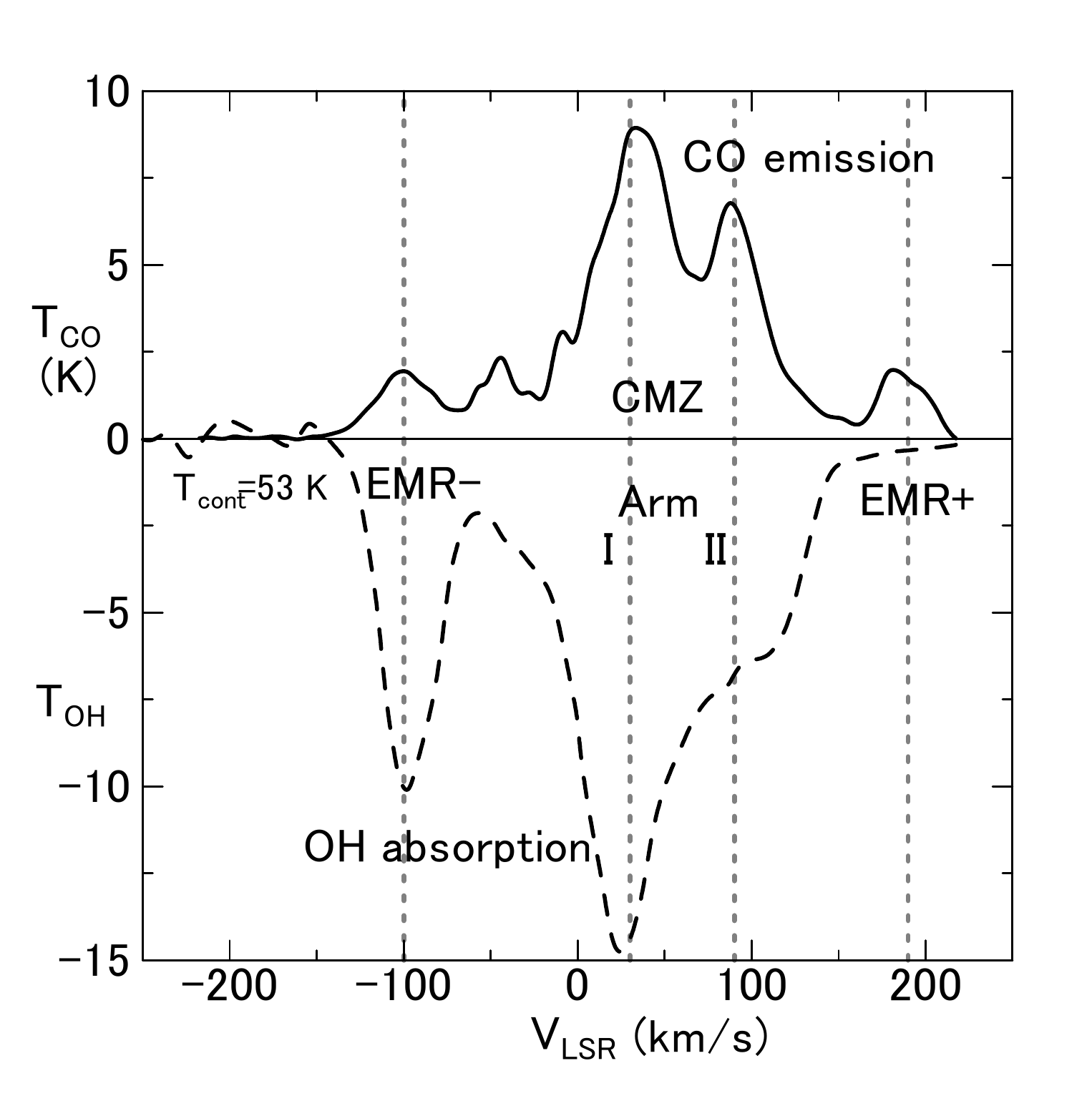}\\
(b)\includegraphics[width=0.7\linewidth]{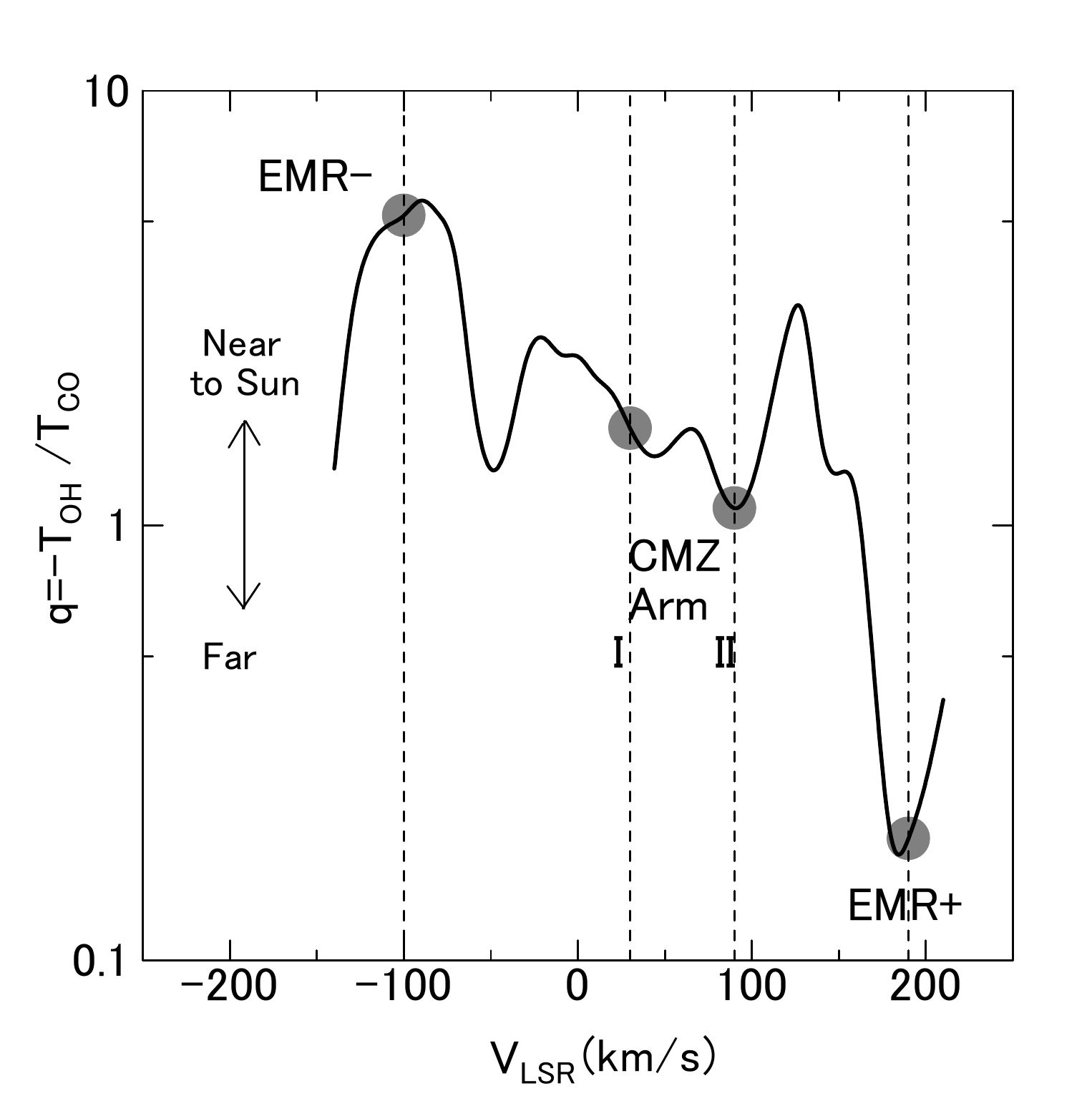}  
\end{center}
\caption{Sawada et al.'s (2004) method to locate molecular gas on the line of sight using the ratio of OH absorption to the CO emission. (a) OH absorption line temperature at $(l,b)=(+0\deg.50, -0\deg.05)$ compared with the CO emission brightness temperature averaged in a circle of diameter $12'$ (same as the beam width of the OH observation). The OH data were taken from Robinson and McKee (1970) and Boyce and Cohen (1994). Vertical dashed lines indicate the velocities for EMR-, EMR+, Arms I and II. (b) OH-to-CO line temperature ratio in logarithmic scaling, showing that the distances are in the order of EMR-, Arm I, Arm II, and EMR+ as marked by the grey circles. 
}
 \label{OHvsCO}  
\end{figure}

Figure \ref{OHvsCO}(a) shows an 18-cm OH line absorption profile at $(l,b)=(+0\deg.50, -0\deg.05)$ from the Jodrell Bank OH survey (Robinson and McKee 1970; Boyce and Cohen 1994), and compare with the CO line profile averaged in a circle of diameter $12'$ (same as the beam width of the OH observation) at the same position. The $(l,\ b)$ direction was so chosen that the Arms I and II are most clearly separated on the LV diagram. The corresponding velocities for EMC at negative and positive velocities, and Arms I and II are indicated by the vertical dashed lines.

The ratio $\tau=T_{\rm OH}/T_{\rm cont}$ indicates an apparent optical depth of the molecular gas in front of the radio continuum background emission (Sawada et al. 2004). The ratio of the OH to CO temperatures, $q=T_{\rm OH}/T_{\rm CO}$, is a measure of the relative depth of the emitting region on the line of sight that the larger is $q$, the nearer is the source. In figure \ref{OHvsCO}(b) we show the variation of $q$ as a function of radial velocity calculated at every 10 \kms interval of $\vlsr$ using Gaussian-running means of the temperatures around each velocity with a half width of 10 \kms. 

This diagram shows that EMR$-$ is located nearest to the observer, and EMR+ is farthest. Arms I and II are located between EMR$-$ and EMR+ with Arm I being slightly closer. This relative location is consistent with the result by Sawada et al. (2004).

\subsection{Parameters of the EMC and CMZ} 
 
Using the observed LV diagrams, we now determine the geometrical parameters of the CMZ and EMC. Obtained parameters are listed in table \ref{para}. In the table, the masses are taken from Sofue (1995a, b) for the conversion factor of $X_{\rm CO}=2.0 \times 10^20$ H$_2$ cm$^{-2}$  (K \kms)$^{-1}$. If we adopt $X_{\rm CO}=1.0 \times 10^20$ H$_2$ cm$^{-2}$  (K \kms)$^{-1}$, the masses, densities and energies are decreased by a factor of 2.

\begin{figure} 
\begin{center} 
(a)\includegraphics[width=0.44\linewidth]{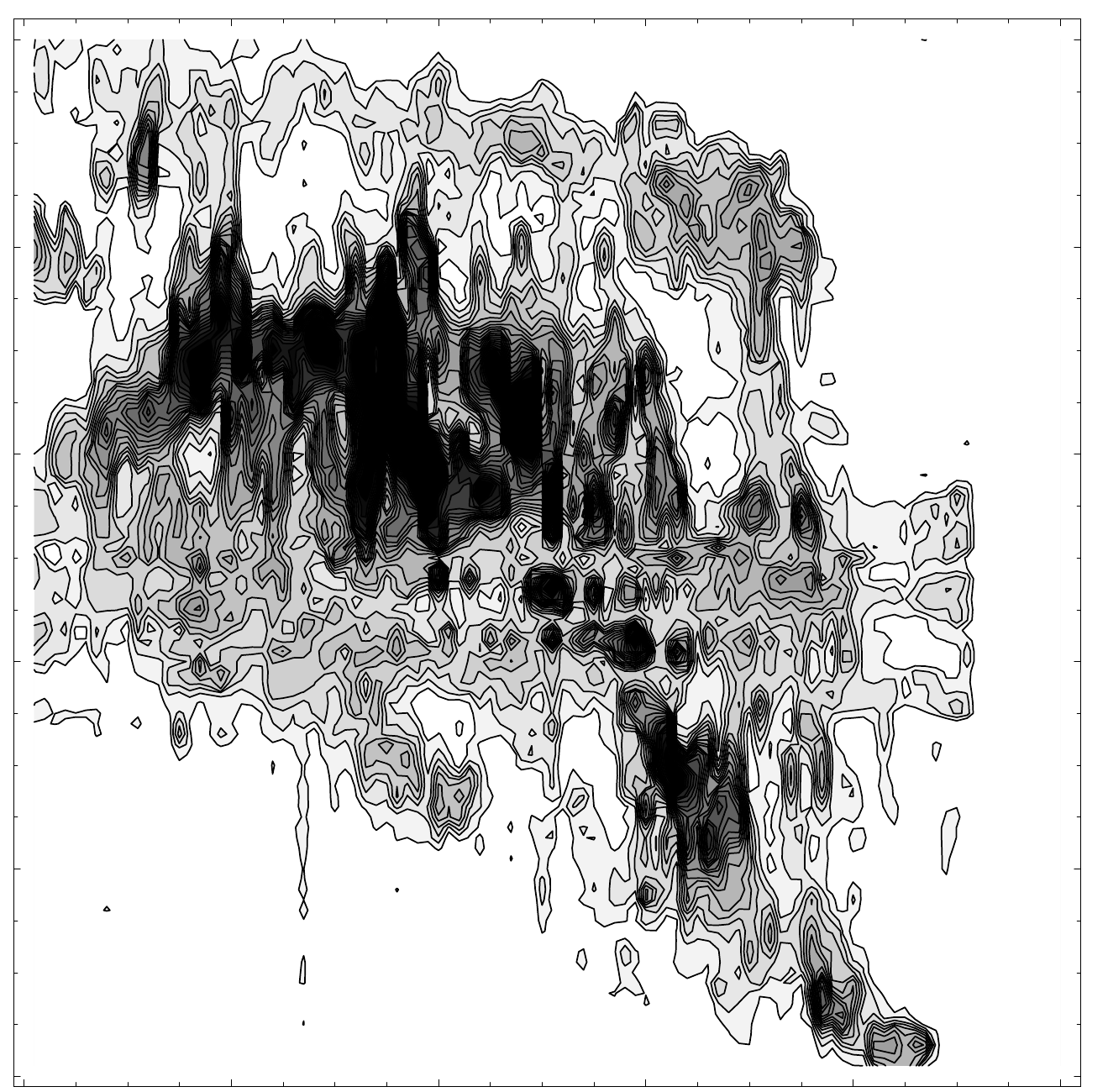} 
\includegraphics[width=0.44\linewidth]{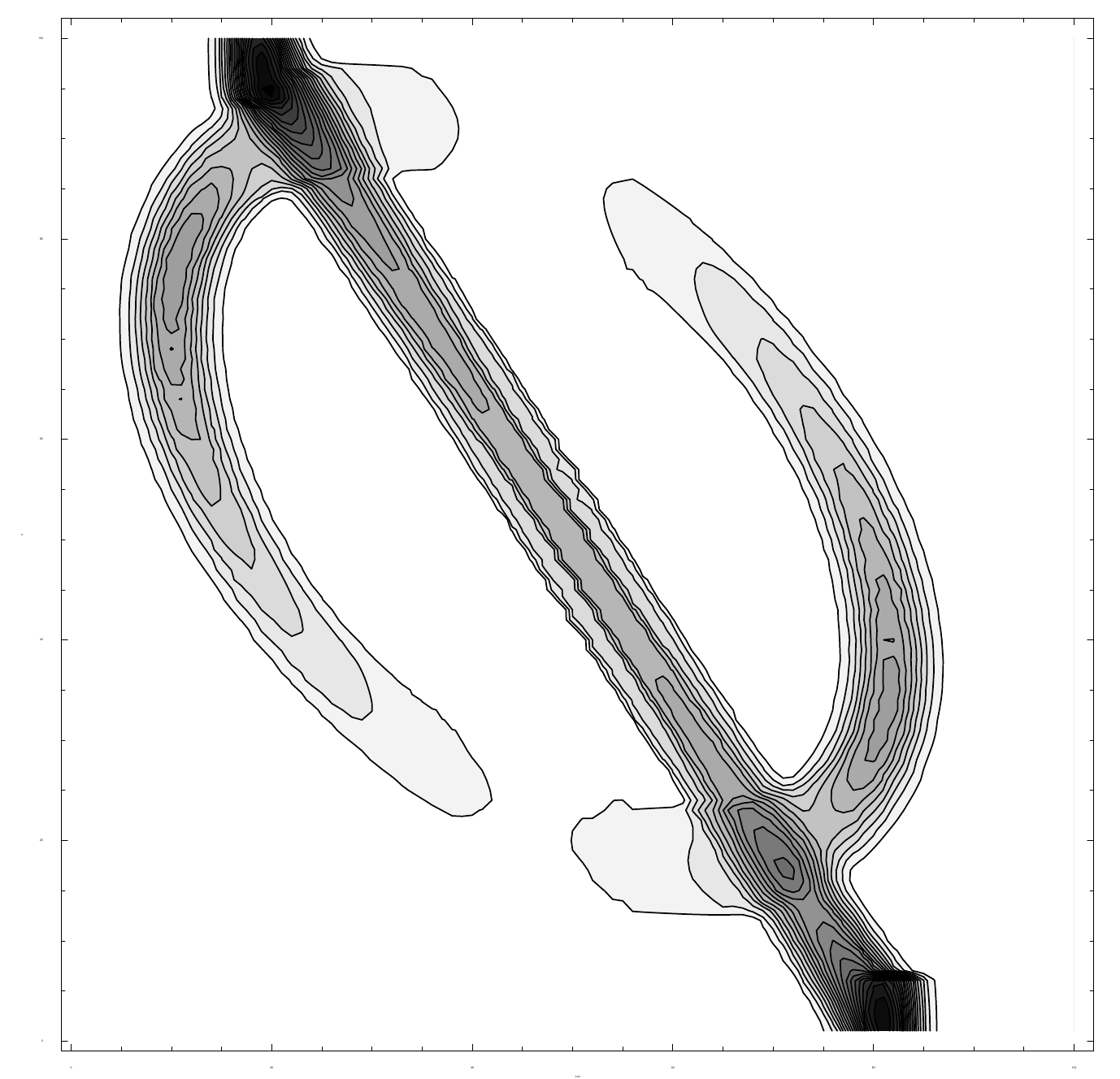}\\ 
(b)\includegraphics[width=0.44\linewidth]{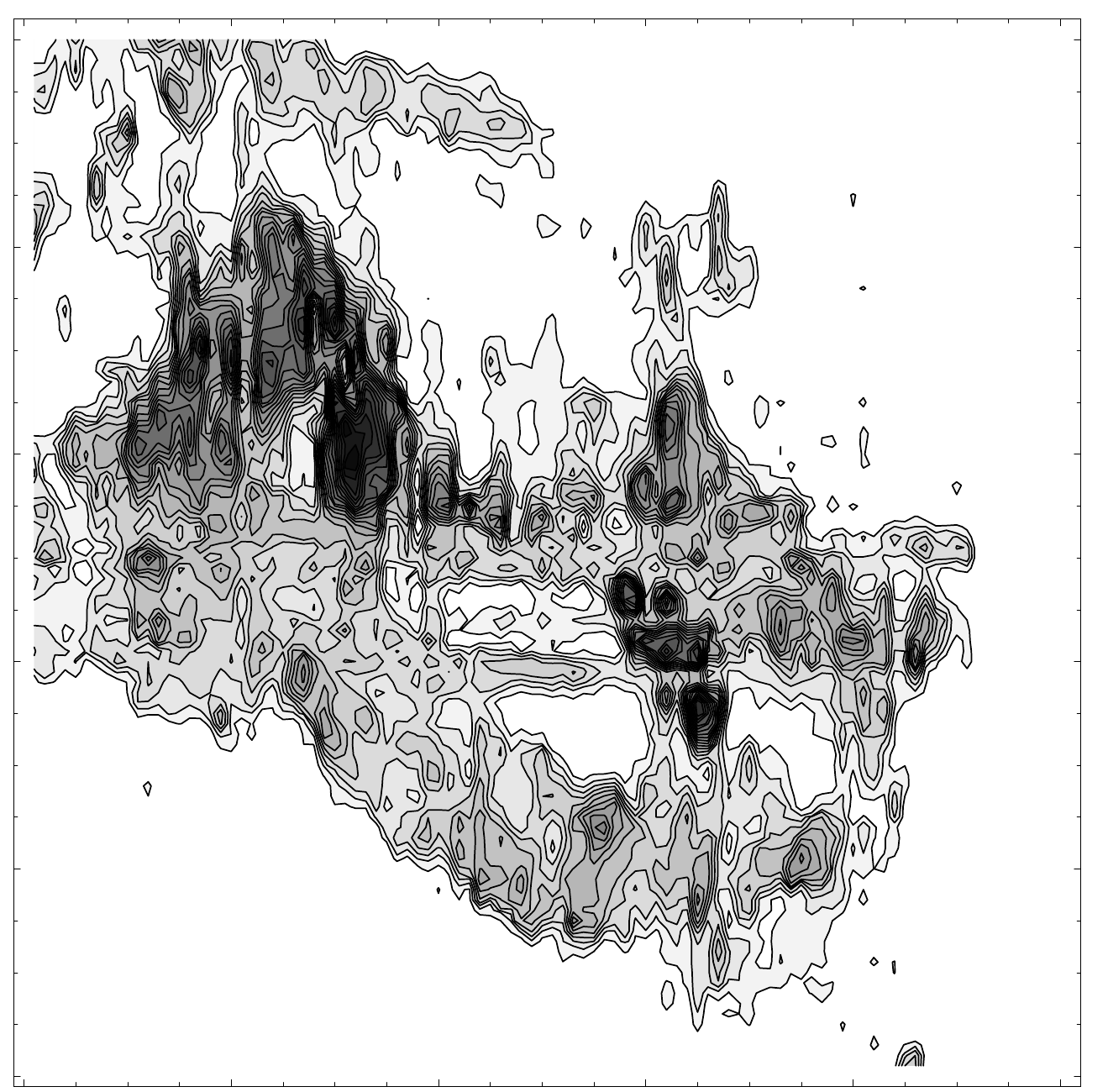} 
\includegraphics[width=0.44\linewidth]{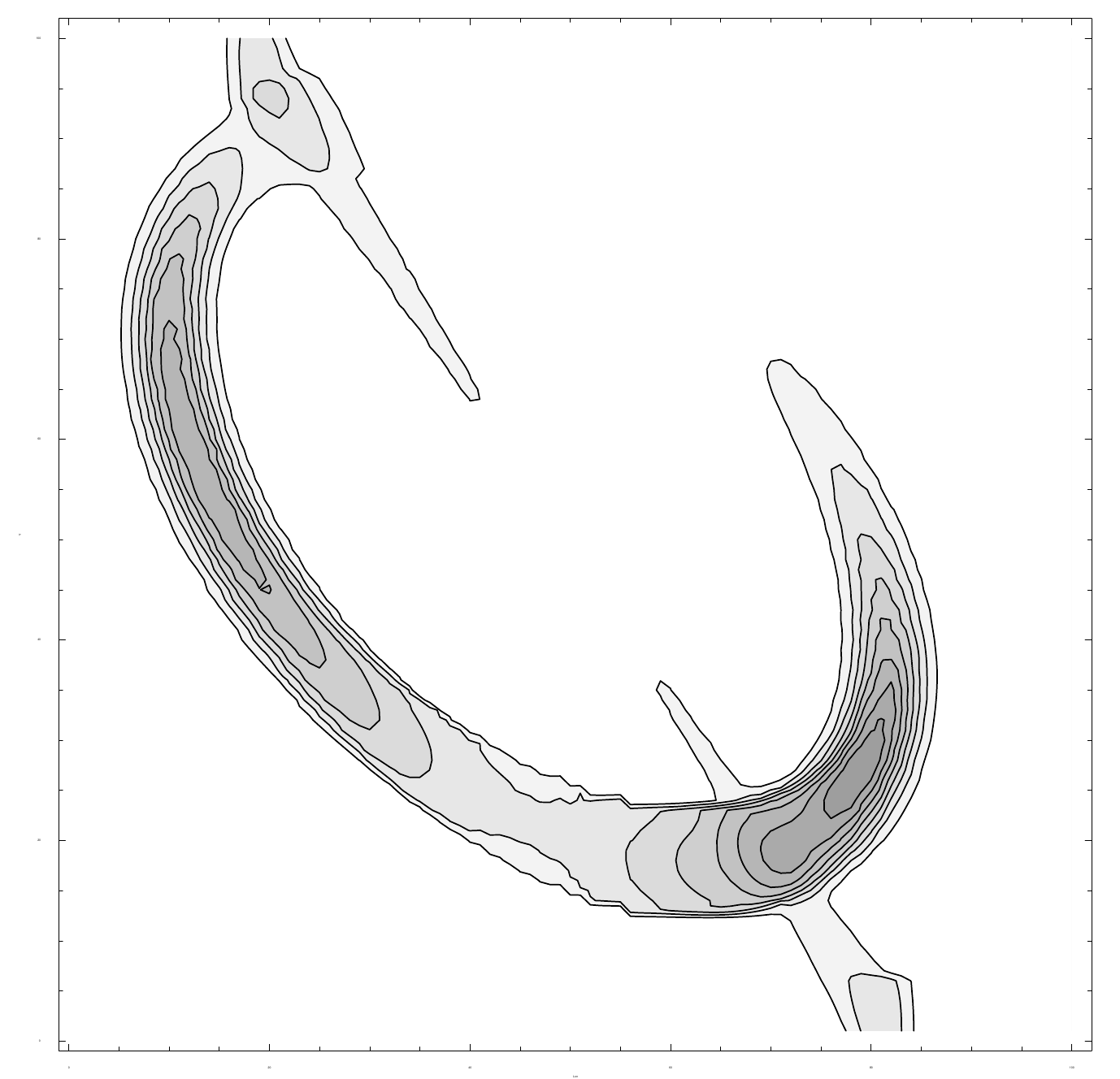}\\ 
(c)\includegraphics[width=0.44\linewidth]{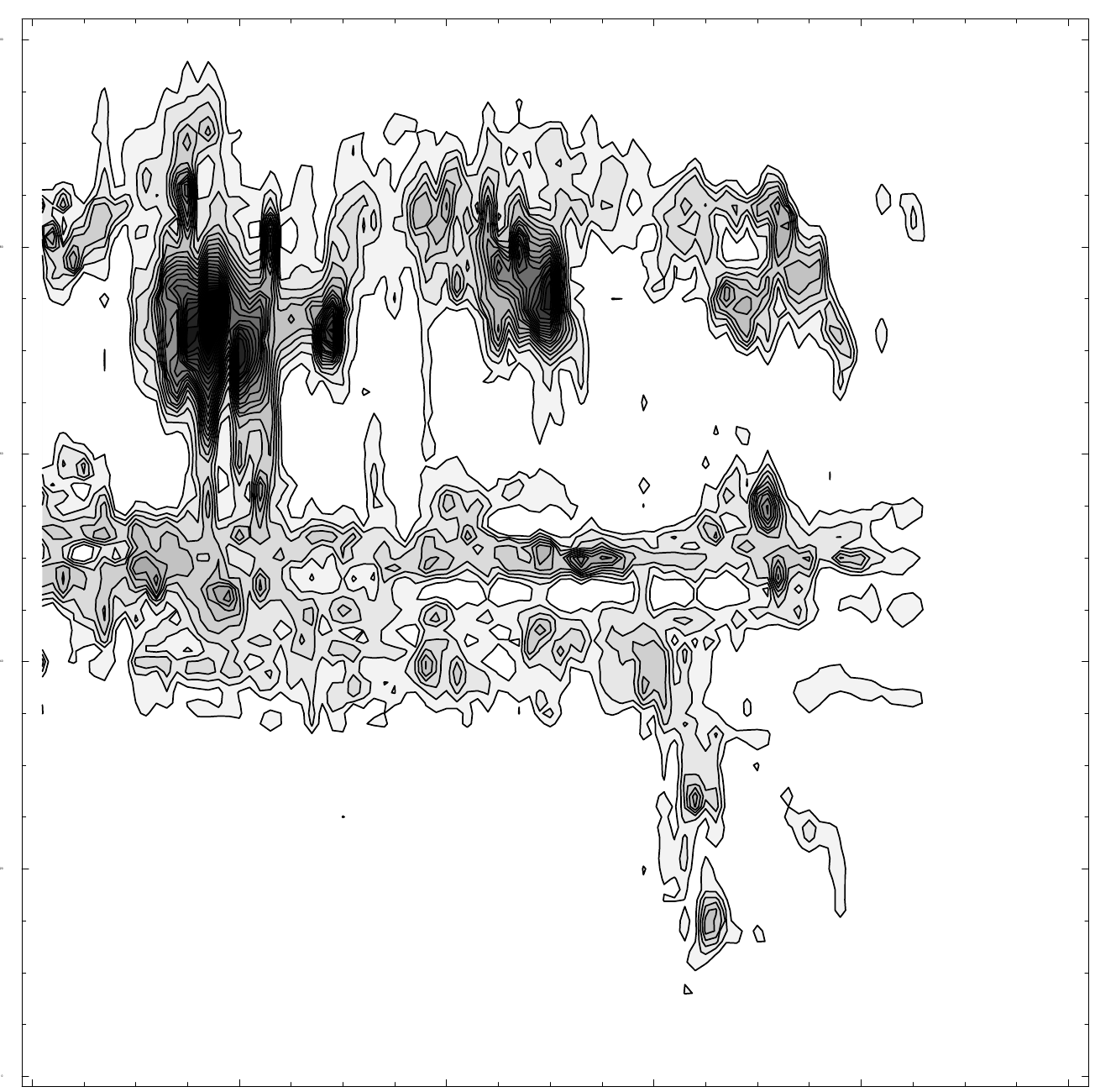} 
\includegraphics[width=0.44\linewidth]{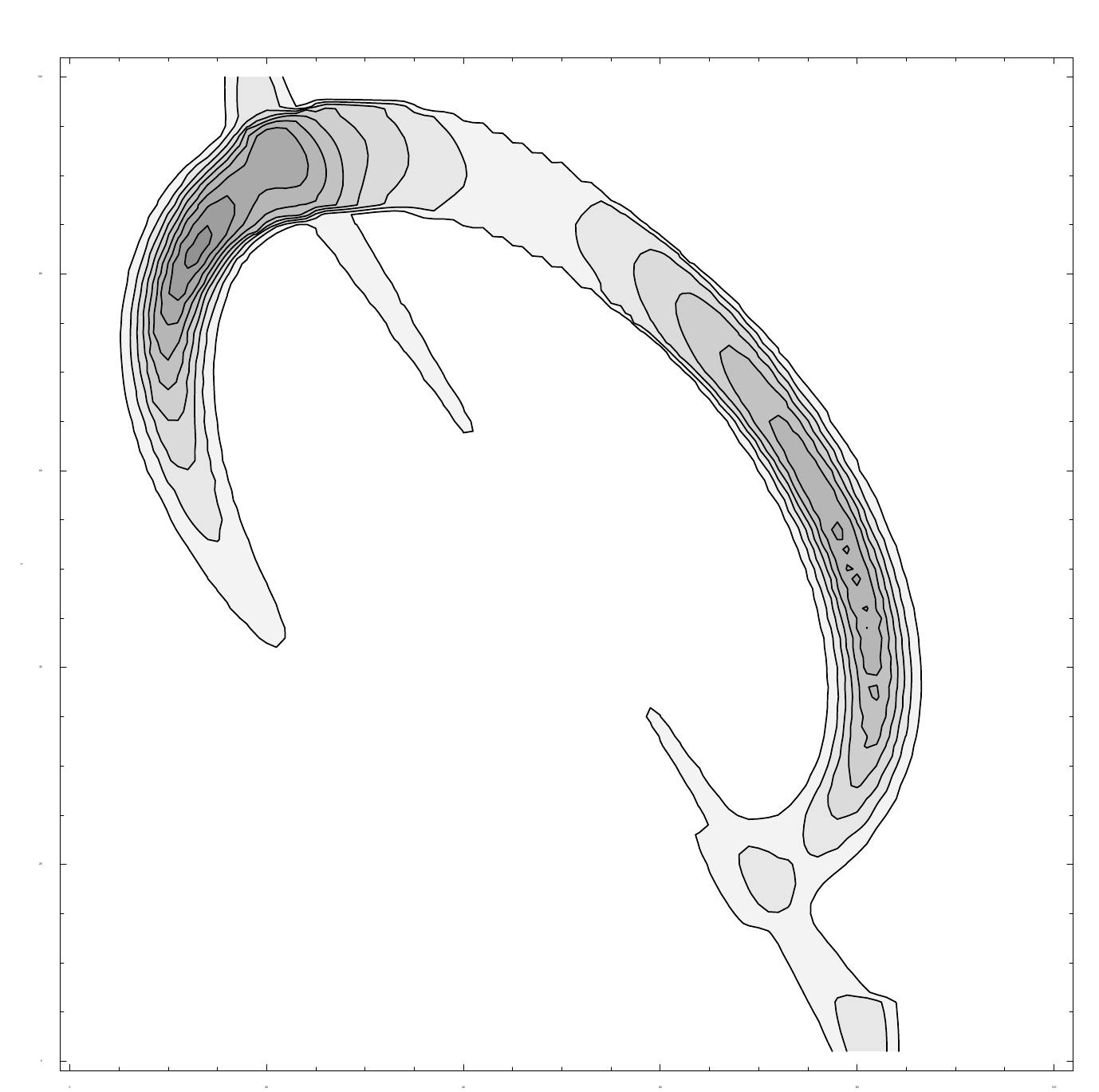}\\ 
(d)\includegraphics[width=0.44\linewidth]{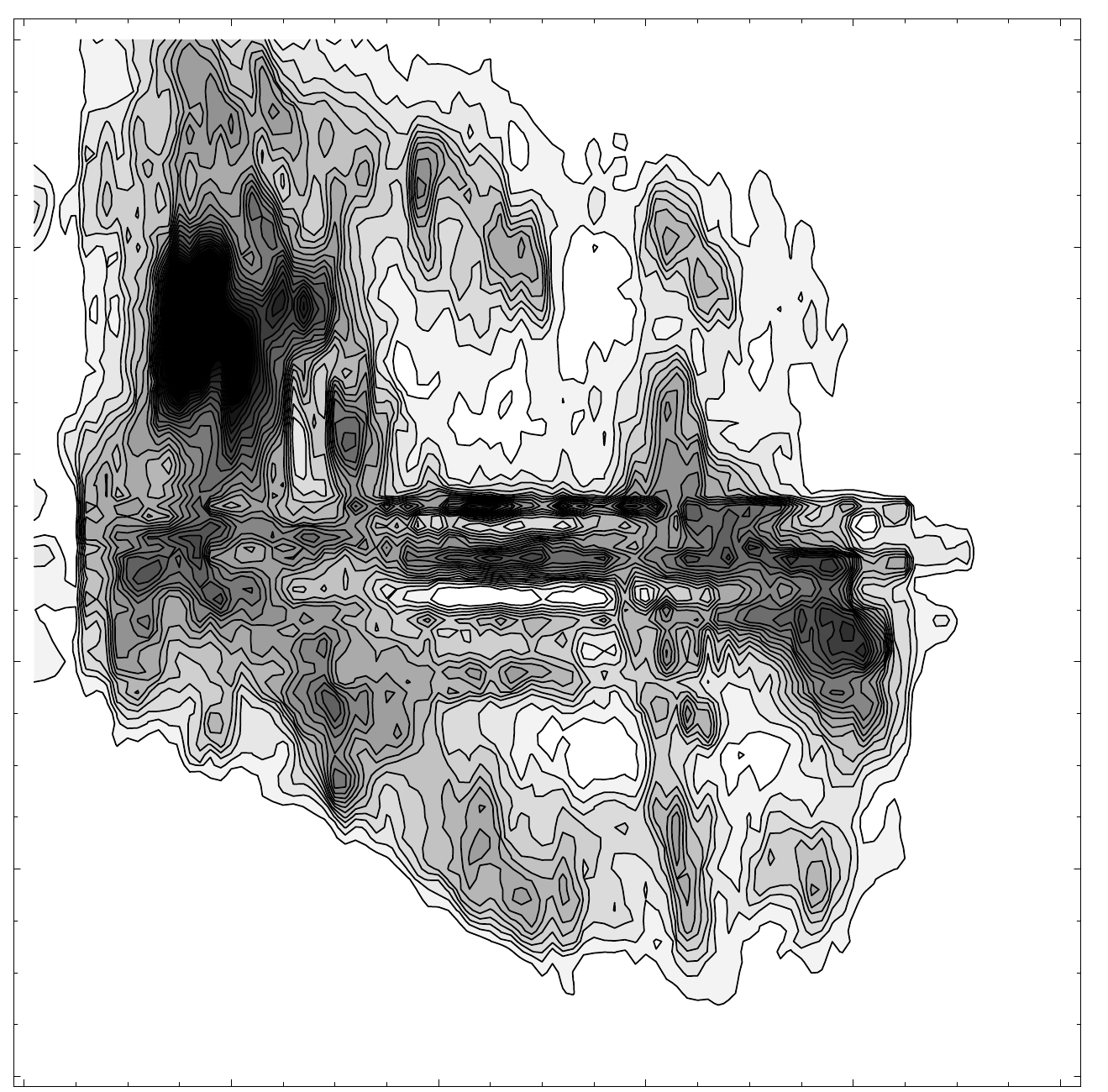} 
\includegraphics[width=0.44\linewidth]{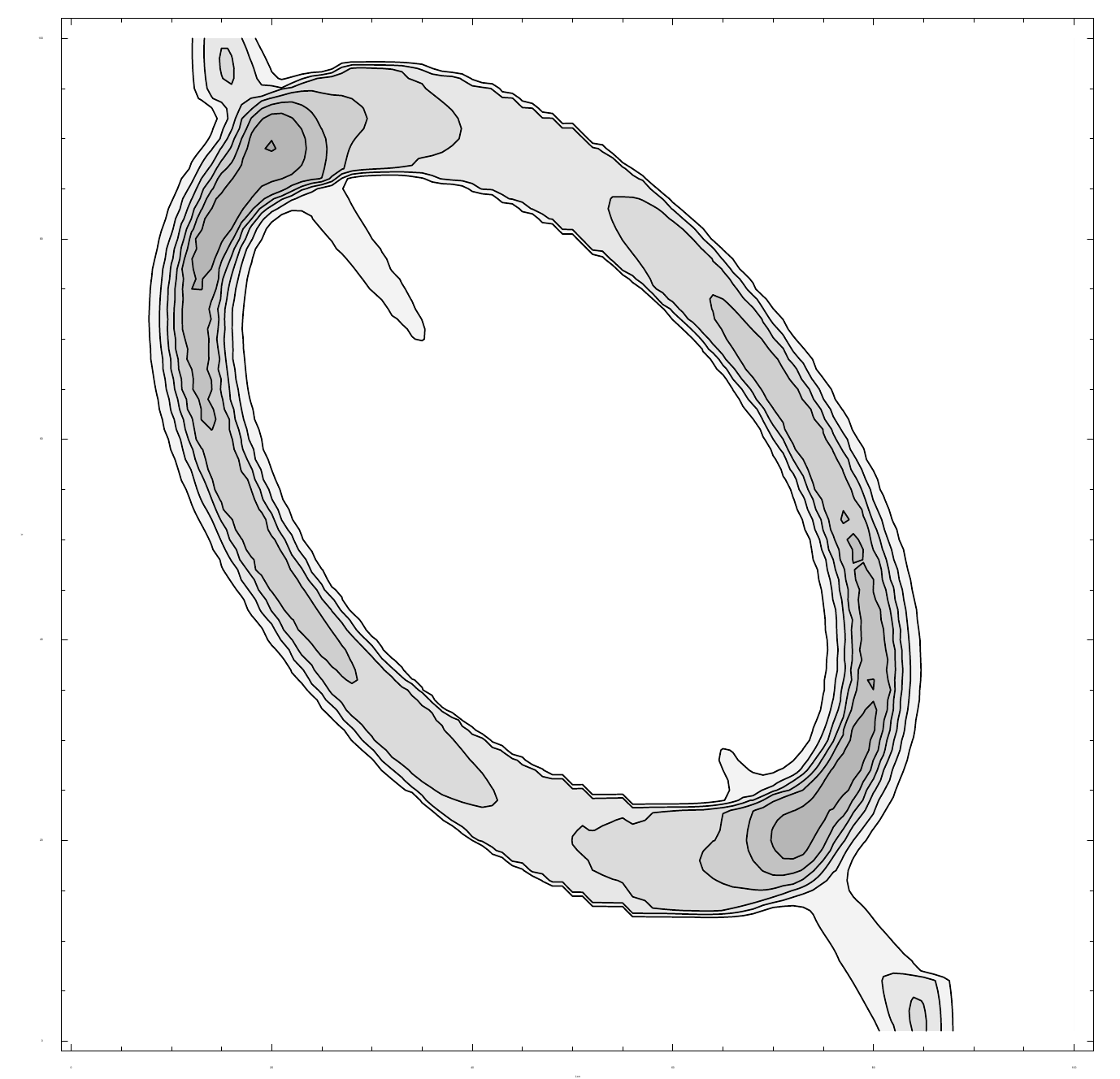}
\end{center}
\caption{Simulated LV diagrams for the parameters listed in table \ref{para} at different latitudes as indicated (right panels), compared with the observed LV diagrams (left panels) at (a) $b=+0\deg.0$, (b) $b=-0\deg.2$, (c) $b=+0\deg.2$, and (d) averaged at $|b|\ge 0\deg.2$. Each plotted area is an LV range at $-2\deg \le l +2\deg$, and $-220\le \vlsr \le 220$ \kms. Contours are drawn every 0.25 K in $\Tb$. }
 \label{LVsimu}  
\end{figure}

In figure \ref{LVsimu}(a) to (c) we show the observed LV diagrams of the CO line emission at different latitudes at $b=0\deg$ and $ \pm 0\deg.2,$ made from the data cube by Oka et al. (1998). The last panel (d) shows an LV diagram averaged at $0\deg.2\le |b|\le 0\deg.6$, which represents the gas distribution out of the galactic plane, avoiding the CMZ gases, and hence represents a purer gas distribution corresponding to the EMC.

The CMZ shows up in the on-plane diagram (a) as the dense tilted LV ridge, while it is hardly seen in the off-plane diagrams from (b) to (d). The EMC is visible as a tilted ellipse in all the figures. The tilted nature is recognized as the asymmetry between the diagrams at positive (b) and negative (c) latitudes. The off-plan diagram in (d) averaged in both latitudes exhibits a more complete LV ellipse.

The CMZ is assumed to represent a rotating ring and the radius and rotation velocity is measured as the end-longitude and velocities on the LV diagram. We measured them to be $\acmz=1\deg.5$ (210 pc) and $\vrot=210$ \kms from the negative longitude edge of the LV ridge. Since the end of the positive-logitude side is not clear, apparently terminated near the dense molecular complex at $(l,\vlsr)\sim (1\deg, +100$ \kms), we adopt these values for negative logitudes.

The EMC is assumed to be composed of an inclined cylinder. The parameters were measured on the LV diagrams, and determined to be $\aemr=1\deg.2$ (160 pc) and $\vrot=75$ \kms. The ellipse centre is located at $(l,\vlsr)=(0\deg.15, +15$ \kms), slightly displaced from the Galactic Center. The measured parameters are listed in table \ref{para}.

The density distributions of the CMZ and EMC is assumed to have the form in the Cartesian coordinates as
\begin{equation}
      \rho=\Sigma_i \rho_i^0 {\rm exp} \left(- \left({r-a_i \over b_i} \right)^2-\left({Z+Y \ {\rm tan}\ i  \over c_i}\right)^2\right),
\end{equation}
where $r=(X^2+Y^2)^{1/2}$ and the radial velocity is related to the rotational and expanding velocities by
\begin{equation}
      v=\vrot X/r+\vexpa Y/r. 
\label{vexp}
\end{equation} 
Here, the $Y$ direction points to the Sun, and $X$ and $Z$ are the perpendicular axes to $y$ with $(X,Y,Z)=(0,0,0)$ denoting the Galactic Center.

      \def\d{\dotfill}
\begin{table}
\caption{Observed parameters of the CMZ and EMC}
\begin{tabular}{l}
\hline\hline 
CMZ  \\
\hline
Radius of the ring, $r$ \d 180 pc \\
Radial full width at half maximum\d 30 pc\\
Vertical full width at half maximum\d 30 pc \\
Rotation velocity, $\vrot$ \d 220 \kms \\
Velocity dispersion \d $\sim 10$ \kms \\
Molecular mass$^\dagger$ \d $\sim 6.1\times 10^7 \Msun$ \\ 
Mean volume density \d $\sim 2.4 \times 10^3\ \Hcc$ \\
      \hline \hline
EMC\\
\hline
Positive-velo. tangent $l$ of LV ellipse \d ~~ $1\deg.5$\\
Negative --- \d $-1\deg.2$ \\ 
Centre longitude \d $0\deg.15$ \\ 
 Radius \d      200 pc      \\
Radial full width at half maximum \d $\sim 40$ pc \\
Vertical full length at half maximum \d $\sim 60$ pc\\
Vertical full length betw. obs. ends\d  $\sim 170$ pc \\  
Tilt angle \d $15\deg$ \\ 
 Molecular gas mass$^\dagger$\d  $ \sim 0.8\times 10^7 \Msun $  \\
 Mean volume density \d $\sim 1.0 \times 10^2\ \Hcc$ \\
 Kinetic energy$^\dagger$\d  $ \sim 2 \times10^{54}$ erg \\
\hline 
Expanding ring fit (explosion model)\\
\hline
Expanding velocity, $\vexpa$ \d 160 \kms \\ 
Rotation velocity $\vrot$ \d 75 \kms \\ 
Centre LSR velocity \d +15 \kms \\ 
\hline
Oval-flow fit (bar model)\\
\hline
Non-circular velocity, $\voval$ \d  160 \kms \\
Phase angle of ellipse\d  $-120\deg$ \\
Axial ratio\d 1.2 \\
\hline \hline
EMC/CMZ ratios\\
\hline
Mass ratio \d 0.13 \\
Density ratio \d 0.04 \\

 \hline
\end{tabular} \\
$\dagger$ Sofue (1995a,b) for conversion factor of $X_{\rm CO}=2.0 \times 10^{20}$ H$_2$ cm$^{-2}$ (K \kms)$^{-1}$. If we adopt $X_{\rm CO}=1.0 \times 10^{20}$ H$_2$ cm$^{-2}$ (K \kms)$^{-1}$, the mass, density, and energy are decreased by a factor of 2. 
\label{para}
\end{table}

The calculated LV diagrams using the measured parameters are shown in the right panels in figure \ref{LVsimu}, which well reproduce the EMR's LV ellipse. The latitudinal asymmetry is recognized in the panels (b) and (c) as due to the inclination of the EMC plane on the line of sight. The oval  continuous LV ridge in the averaged LV diagram at $|b|\ge 0\deg.2$ is almost perfectly fitted by the model.

On the other hand, the CMZ is not well reproduced. The negative-longitude side of the LV ridge of the CMZ is well reproduced by the calculation, but the positive-logitude side cannot be reproduced. Namely, the observed LV ridge in the galactic plane ($b=0\deg$) does not continue to the maximum rotation speed expected by the rotating ring or arms, but is terminated by the dense extended gas component at $(l,\vlsr)\sim (1\deg, 100$ \kms). Such termination of the LV ridge could be understood, if the ring or the arms are not a continuous loop structure, but is lacking its significant portion at the positive-longitude side at $l\sim 0\deg.7$.  
  
The LV feature of the EMR may also be approximated by an oval flow by 
\begin{equation}
\vlsr \simeq \voval (Y/r \ {\rm sin} \ \theta -X/r \ {\rm cos} \ \theta),
\label{vplg}
\end{equation}
wher the ellipse radius is expressed by an ellipse of oblateness $\epsilon$ around the circle as
\begin{equation}
a_{\rm oval} \simeq a_{\rm EMR} (1+\epsilon {\rm sin} \ (\theta-90\deg))/(1+\epsilon).
\label{vplg}
\end{equation}
Here, $\voval$ represents the flow velocity and $\theta$ is the phase angle of the motion. By eye fitting, the observed LV oval is approximately reproduced by the parameters as $\theta \simeq -120\deg$, $\voval \simeq 160$ \kms, and $\epsilon \simeq 0.2$. It is thus shown that the EMR and bar models are nearly identical in so far as the LV behavior is concerned. 

\subsection{3D view}

\begin{figure} 
\begin{center} 
\includegraphics[width=0.8\linewidth]{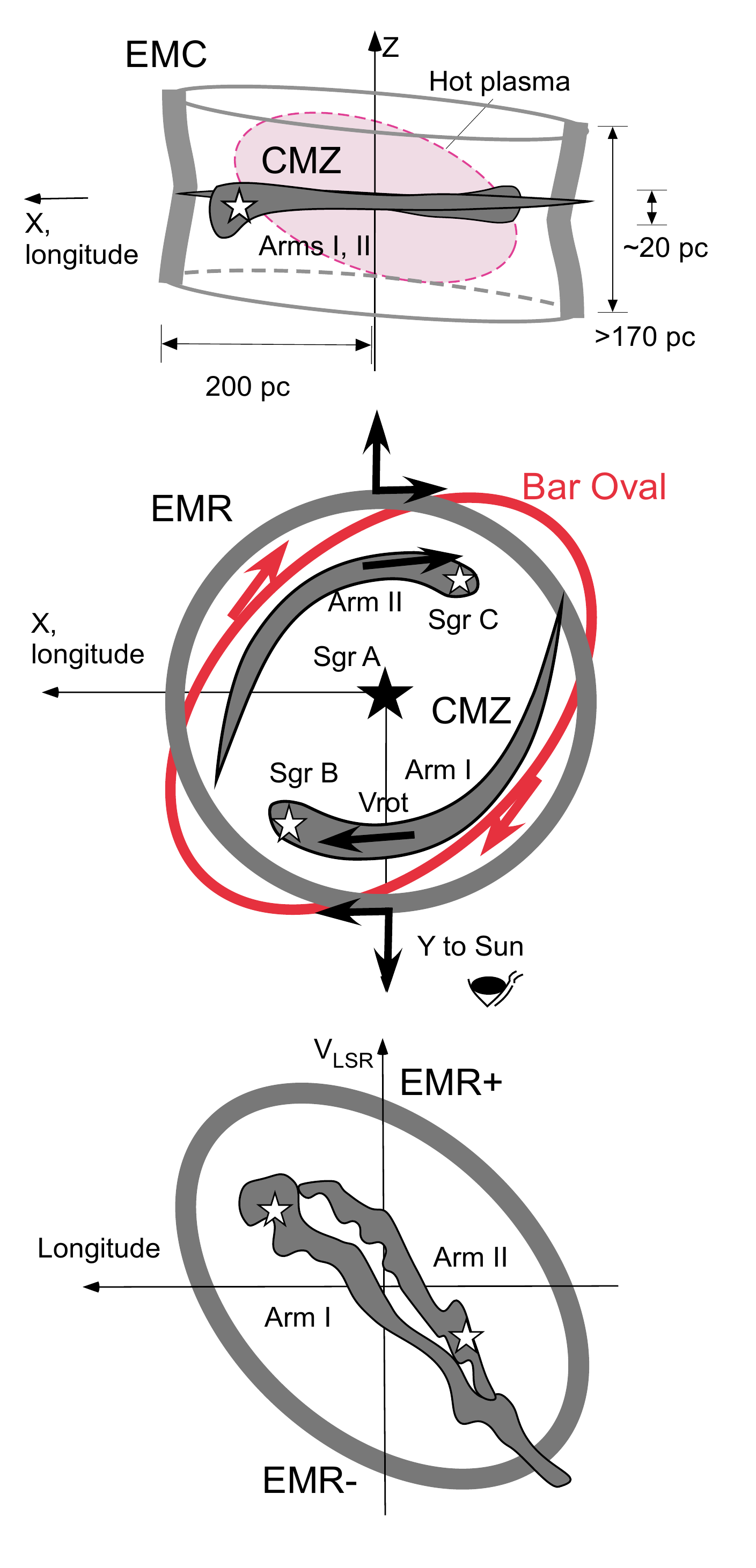} 
\end{center}
\caption{Schematic 3D view of the CMZ and EMR. [Top] $X-Z$ projection, showing the large EMC extent for more than $\sim 170$ pc in the $z$ direction. Hot plasma observed by Uchiyama et al. (1991) is indicated by the ellipse. [Middle] $X-Y$ projection, showing the spatial relationship of the EMC and Arms I and II of the CMZ. The red ellipse illustrates the bar-induced oval orbit. [Bottom] Schematic LV diagram with Arms I and II from Sofue (1995a).}
 \label{3Dview}  
\end{figure}

From the cylindrical structure of the EMR/EMC and the above consideration of the line-of-sight distribution of the molecular gas, we may draw a possible 3D configuration of the molecular structures in the Galactic Centre as illustrated in figure \ref{3Dview}. We may also emphasize that the main body of the GC gas is confined in the CMZ with total mass of $\sim 3-8 \times 10^7 \Msun$ (Tsuboi et al. 1999), whereas the EMR shares only a small portion of the molecular mass ($\sim 8\times 10^6 \Msun$).

\section{Discussion}
 
\subsection{Models for understanding the 3D EMC and CMZ}
 
\subsubsection{Bar model} 
 The noncircular motions observed in the GC have been modeled by streaming flows in a bar, and the LV behavior was named parallelogram after the shape of the lowest contours (Binney et al. 1991, 1997). Accumulated evidences of a bar in the Milky Way (Blitz and Spergel 1991a, b; Dwek et al. 1995) stimulated hydrodynamical simulations of LV behaviors in a bar potential (Englmaier and Gerhard 1999; Fux 1999; Mulder and Liem 1986; Rodriguez-Fernandez and Combes 2008; Baba et al. 2010; Ridley et al. 2017; Shin et al. 2017). 
 
 Accordingly, many of the galactic rings and/or arms such as the 4-kpc molecular ring, 3-kpc expanding ring, the EMR, and CMZ have been believed to be the product of the bar dynamics. Advantage of the bar model is that it explains the non-circular motion without assuming any violent phenomenon.

  However, the current simulations do not necessarily explain the observed 3D complexity about the CMZ and EMR: (a) different vertical exents, (b) different LV behaviors (paralellogram in the plane vs oval in off-plane region), and (c) the large contrast for an order of magnitude between the densities as well as the masses. In order to simulate these complexity inside $\sim 2\deg$ of the GC, more sophysiticated 3D models with much higher resolution would be necessary

Bar accretion is one of the major concerns for triggering starburst and central activity in galaxies (Roberts et al. 1979; Kenney et al. 1993; Jogee et al. 2005; Koda and Sofue 2006; Hsieh et al. 2011; Salak et al. 2016). If the Galaxy is barred, the gas accretion results in energy injection of $\sim 10^{55}$ erg at the centre thanks to the release of nuclear energy via starburst and gravitational energy of the SMBH. Evidences for such activities in the GC have been accumulating as described in the next subsection by (i) to (v).

\subsubsection{Explosion model}  

\begin{figure} 
\begin{center}   
\includegraphics[width=0.49\linewidth]{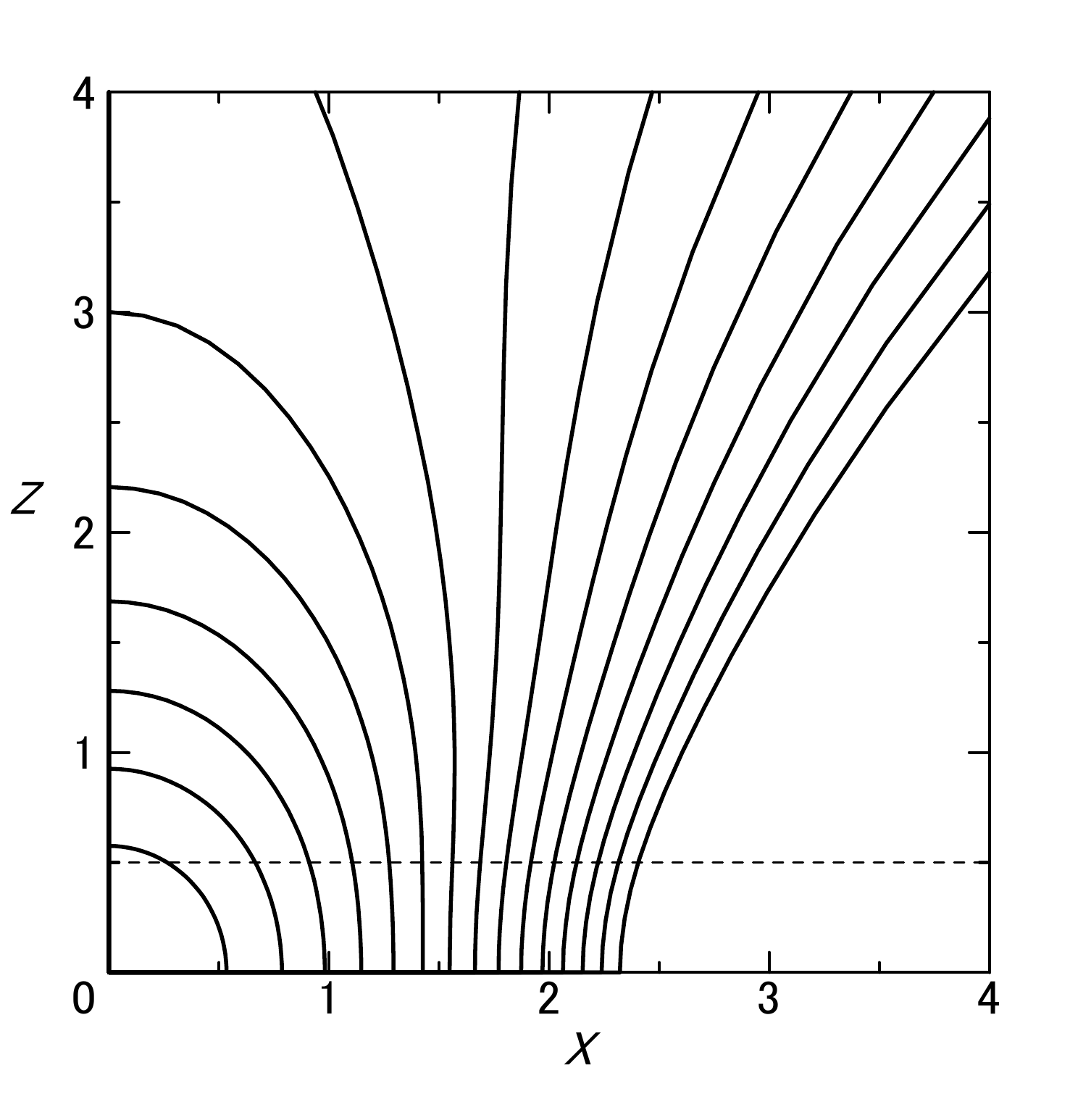}
\includegraphics[width=0.49\linewidth]{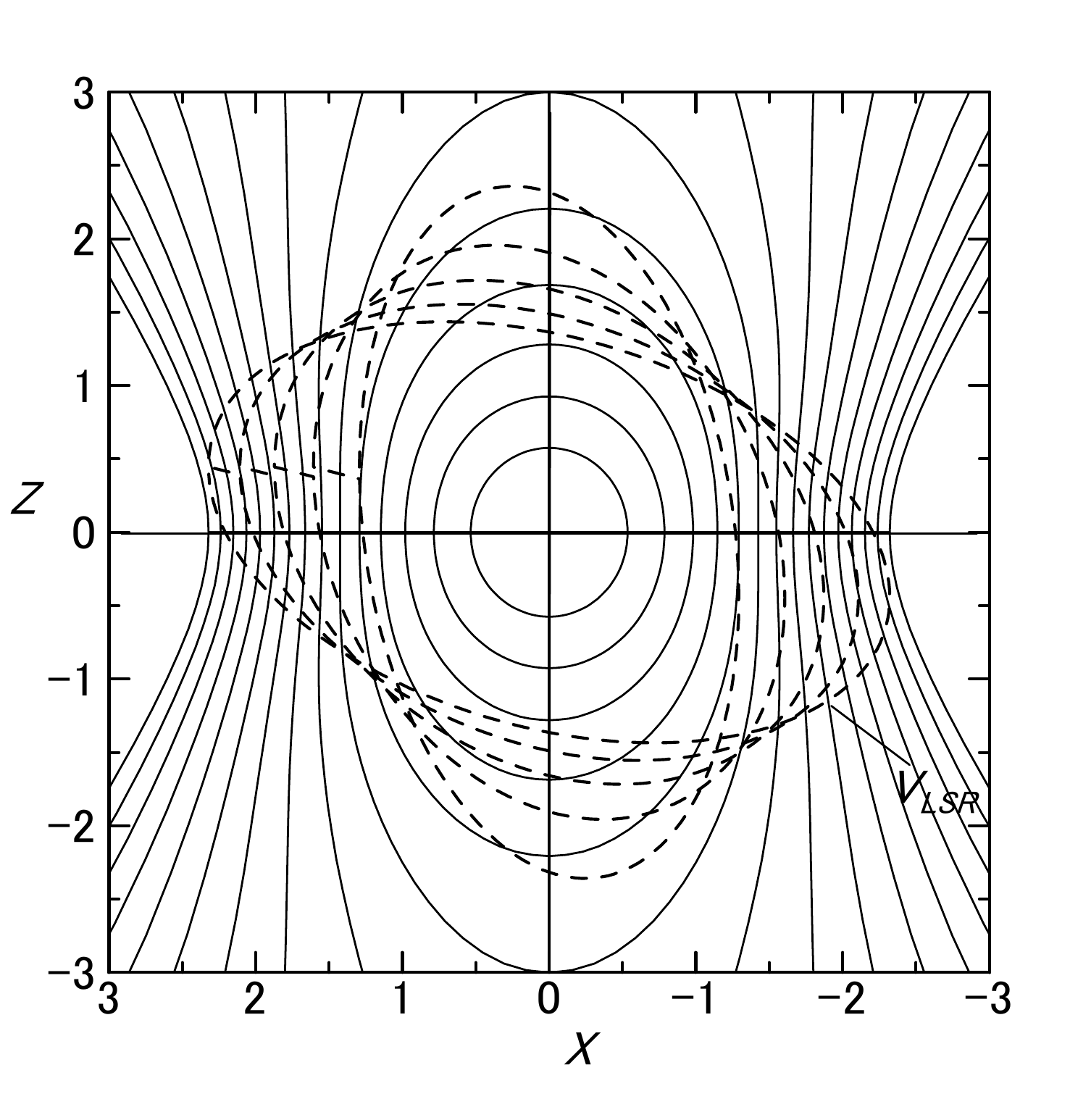}\\  
(a)~~~~~~~~~~~~~~~~~~~~~~~~~~~~~~~~~~~~~~~~~~(b)\\ 
\end{center}
\caption{
(a) Propagation of a shock front in the $(X, Z)$ plane (in unit of $100$ pc) in a one-layered sech$^2$ disc with $\rho_0=20$ H cm$^{-3}$ and $h=50$ pc at every $6\times 10^4$ y up to $9\times 10^5$ y for injected energy of $E_0=1.8\times 10^{54}$ erg. The horizontal dashed line shows the scale height $h=50$ pc of the disc. 
(b) Same as (a), showing the four quadrants, mimicking the EMC. Dashed ellipses show LV diagrams in the galactic plane with arbitrary velocity scale. 
}
 \label{shock}  
\end{figure}

In order to examine if the EMC is explained by explosion model, we calculated the propagation of a shock wave from the GC using the Sakashita's (1971) method (see also Moellenhof 1976; Sofue 2000). The unperturbed gas is assumed to be distributed in a hydrostatic layer (Spitzer 1942) embedded in a halo of uniform density. The disc-center density is assumed to be $\rho_0$=20 H cm$^{-3}$, the scale height $h=50$ pc, the halo density $\rho_{\rm halo}=10^{-5}$ H cm$^{^3}$, and the injected energy of $E_0=1/2 M_0 V_0^2 \sim 1.8\times 10^{54}$ erg, which is equal to thermal energy of hot plasma with sound velocity $V_0 = 300$ \kms in a sphere of radius $h$ of $M_0=4 \pi/3 h^3 \rho_0 = 2\times 10^6\Msun$. Figure \ref{shock} shows the calculated result. 

The shock wave expands faster in the $Z$ direction, and blows off into the halo due to the steep pressure gradient in the $Z$ direction. The near-plane shock is decelerated by the high-density disc, and the front shape becomes cylindrical at $t\sim 2$ at $X\sim 200$ pc. The rotation speed decreases due to the conservation of angular momentum.  

As to the energy source of the explosive event we may consider several observational facts in the GC.

{\bf (i)  Hot plasma}: The hot plasma of $ \sim 10^8$ K in 6.7 keV iron line containing thermal energy of $ \sim 3 \times 10^{53}$ erg (Koyama et al. 1989, 1996; Yamauchi et al. 1993) evidences an explosive phenomenon in the GC. The gas is extended for $\sim 1\deg.8 \times 0\deg.5 \ (\pm 0\deg.9\times \pm 0\deg.25)$ along the galactic plane about the GC (Yamauchi et al. 1993), coincident with the interior of the EMC.  

{\bf (ii) GC Lobe}: The GCL of a size $\sim 150\times 200$ pc is thought to be a bubble driven by energy injection in the GC of the order of $\sim 10^ {54}$ erg (Sofue and Handa 1984). However, its shape, size, and estimated time scale are not coincident with the EMC, suggesting that it may not be directly related to EMC, but is a phenomenon after the EMC.

{\bf (iii) Fermi Bubbles}: The $\gamma$-ray Fermi Bubbles are considered to be driven by an event of the order of $\sim 10^{55}$ ergs, and  their time scale is several $10^6$ y (Su et al 2010). The time scale suggests the relation to the EMC. In fact, the shock wave model predicts that shock front breaks into the halo and form a dumbbell-shaped bubbles in the later stage. 

\begin{figure} 
\begin{center} 
\includegraphics[width=0.8\linewidth]{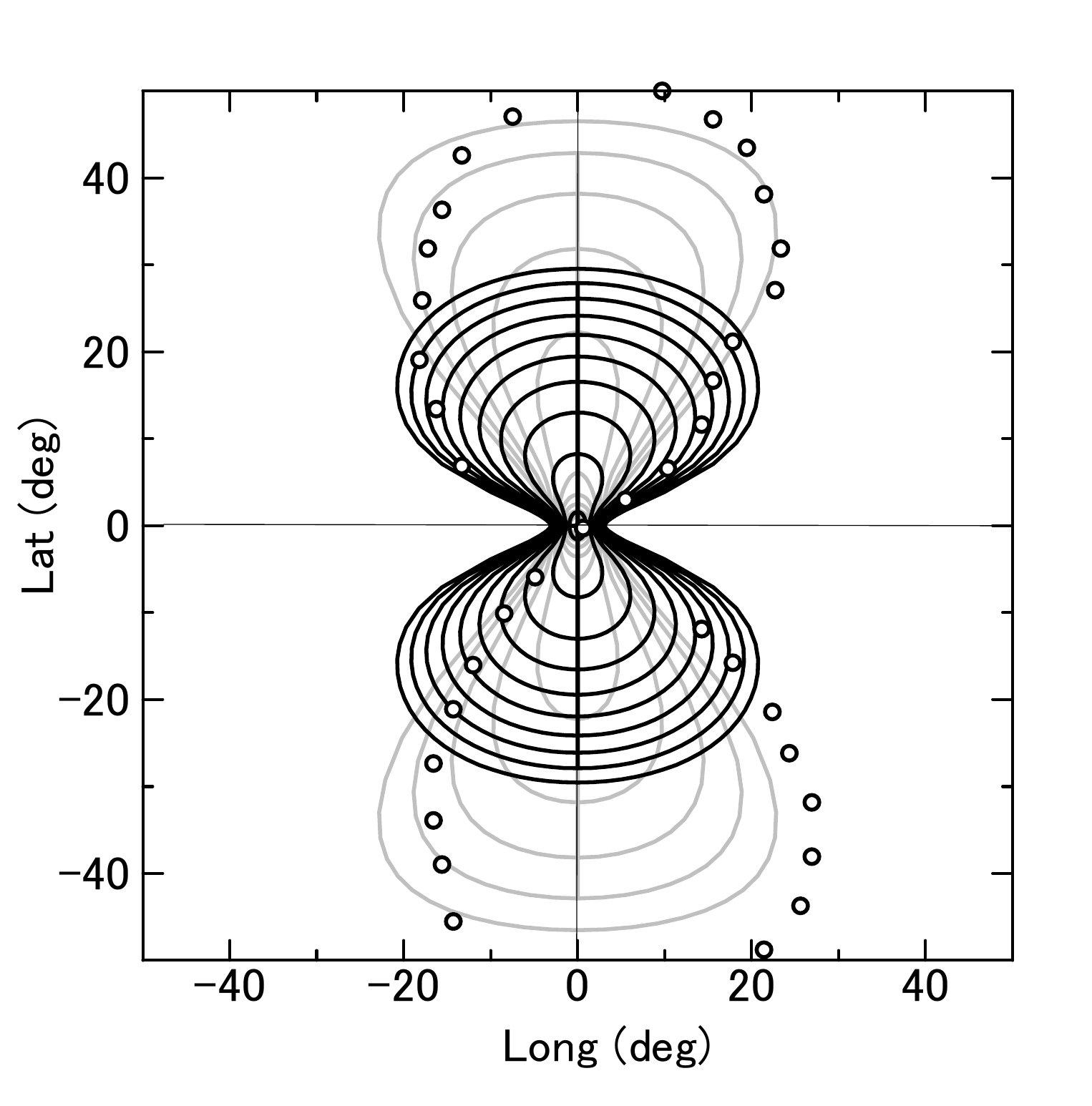} \\ 
\end{center}
\caption{Shock front up to $t=10$ ($\simeq 3\times 10^6$ y) cut at $Y=0$ projected on the sky. Open dots indicate the shape of Fermi Bubbles (Su et al. 2010). Grey lines show a toy model to fit the FB, taking $h=100$ pc and an almost empty halo.  }
 \label{bhs}  
\end{figure}

Figure \ref{bhs} shows the shock propagation up to $3\times 10^6$ y compared with the FB envelope. In the figure we also show a toy model in order to better fit the FB for the same paramters except for a larger scale height of $h=200$ pc and an almost empty halo.

{\bf (iv) Bipolar Hyper Shells (BHS)}: The BHS including the North Polar Spur (Sofue 2000; Sofue et al. 2016) are considered to be the result of an explosion or a starburst of energy $\sim 10^{55}$ erg occurred $\sim 10^7$ y ago. The event may not be directly related to the EMC. However, it indicates recurrent explosions in the past, which may have resulted in outer rings such as the 130-\kms arms and 3-kpc ring (Sofue 1976, 1984).
  
 (v) {\bf Gas supply by a bar}: An indirect, but the ultimate evidence for the starburst and nuclear activity in the GC is the high rate of accretion of disc gas by the bar. The events occur intermittently, depending on how the bar-swept empty disc is fueled again.  
 
\subsection{Concluding remarks}
 
We presented a 3D view of the central 200 pc of the Galactic Centre based on the high-resolution and sensitivity data of $^{12}$CO $(J=1-0)$-line observations with the Nobeyama 45-m telescope (Oka et al. 1998).

The EMR (EMC) and CMZ exhibit quite different distributions and kinematics. The EMR composes a vertical cylinder with the toal length of $\sim 170$ pc having large non-circular velocities, while the CMZ is tightly distributed close the galactic plane in a dense ring/arms in nearly rigid-body rotation. It is also stressed that the total mass and density of the EMR are an order of magnitude smaller than those of the CMZ.

In order to understand the 3D view, we considered both the bar and explosion models. Highlighting the explosion model, we calculated the shock wave propagation from the centre, which approximately reproduced the observed cylindrical structure of the EMC. 

We argued that the bar and central activity, both evidenced observationally in the Galaxy, are deeply coupled. Mutual feedback, how the bar induces explosion via the gas accretion, and how the explosion disturbs the gas disc, would be an intriguing subject for the future in hope of discriminating or integrating the models.  

\vskip 5mm

\noindent {\it Acknowledgements}: The author is indebted to Prof. Tomoharu Oka of Keio University for providing the CO line survey data using the Nobeyama 45-m telescope and for valuable comments.

\end{document}